\documentclass[a4,center,fleqn]{article}

\usepackage{algorithm}
\usepackage{algorithmicx}
\usepackage{algpseudocode}
\usepackage{amsmath}
\usepackage{amsthm}

\usepackage{xcolor}
\usepackage{xspace}
\usepackage{color}
\usepackage{caption}
\usepackage[font=scriptsize]{subcaption}

\usepackage{lscape}
\usepackage{longtable}
\usepackage{paralist}
\usepackage{multirow}
\usepackage{tabularx}
\usepackage{booktabs}
\usepackage{array}
\usepackage{rotating}
\usepackage{colortbl}
\usepackage{url}

\usepackage{authblk}

\newcommand\subcolbeg{\setlength{\extrarowheight}{.0ex}\renewcommand{\tabcolsep}{1pt}\tiny}
\newcommand\subcolend{\setlength{\extrarowheight}{.4ex}}
\newcommand\subcolvspace{\vspace{.02ex}}
\newcommand{\Report}{\textbf{report }}

\newcommand{\obacht}[2][\empty]{{\color{cyan} \textsl{\scriptsize {(\ifthenelse{\equal{#1}{\empty}}{}{#1: }{#2})}}}}
\newcommand{\answer}[2][\empty]{{\color{orange} \textsl{\scriptsize {(\ifthenelse{\equal{#1}{\empty}}{}{#1: }{#2})}}}}

\newcommand{\eg}{{e.g.}\xspace}
\newcommand{\ie}{{i.e.}\xspace}
\newcommand{\CC}{C\raise.06ex\hbox{\tt ++}\xspace}

\begin{document}

\title{Fast and sensitive read mapping with approximate seeds and multiple backtracking}

\author{Enrico Siragusa}
\author{David Weese}
\author{Knut Reinert}

\affil{Department of Mathematics and Computer Science, Freie Universit\"at Berlin}

\maketitle

\begin{abstract}
We present \emph{Masai}, a read mapper representing the state of the art in terms of speed and sensitivity.
\textcolor{black}{Our tool is an order of magnitude faster than RazerS\,3 and mrFAST, 2--3 times faster and more accurate than Bowtie\,2 and BWA.}
The novelties of our read mapper are filtration with approximate seeds and a method for multiple backtracking.
Approximate seeds, compared to exact seeds, increase filtration specificity while preserving sensitivity.
Multiple backtracking amortizes the cost of searching a large set of seeds by taking advantage of the repetitiveness of next-generation sequencing data.
Combined together, these two methods significantly speed up approximate search on genomic datasets.
Masai is implemented in \CC using the SeqAn library.
The source code is distributed under the BSD license and binaries for Linux, Mac OS X and Windows can be freely downloaded from \url{http://www.seqan.de/projects/masai}.
\end{abstract}


\section{Introduction}

Next-generation sequencing (NGS) allows to produce billions of base pairs (bp) within days in the form of reads of length 100\,bp and more.
It is an invaluable technology for a multitude of applications in biomedicine, \eg detection of SNPs and large genomic variations, targeted or de-novo genome or transcriptome assembly, isoform prediction and quantification, identification of transcription factor binding sites or methylation patterns.
In many of these applications mapping sequenced reads to their potential origin in a reference genome is the first fundamental step preceding downstream analyses.

Because of sequencing errors and \textcolor{black}{genomic} variations not all reads occur exactly in a reference genome.
Therefore approximate occurrences must be considered and algorithms for approximate string matching tolerating mismatches and indels must be applied to solve the problem.
Furthermore, because of homologous and low complexity regions not all reads occur uniquely in a reference genome.
Therefore in some applications, \eg CNVs calling, all approximate occurrences which could be potential origins must be considered.

\subsection{Previous work}

All current read mappers can be broadly classified as \emph{best-mappers} or \emph{all-mappers}.
Tools in the first class aim at finding the best mapping location for a read according to a scoring scheme eventually taking base quality values into account, while those in the second class aim at enumerating a comprehensive set of locations.

Most prominent best-mappers are based on \emph{backtracking algorithms} for approximate string matching \textcolor{black}{\cite{Navarro2001}}.
Substrings of the reference genome within an absolute number of errors from a read \textcolor{black}{are} recursively enumerated using a suffix or prefix tree of the reference genome.
Since the time complexity of backtracking grows exponentially with the absolute number of errors considered, this method alone \textcolor{black}{is} impractical when mapping \textcolor{black}{whole} reads with moderate error rates.
Hence popular best-mappers \cite{Bowtie2, BWA, Soap2} apply heuristics to reduce and prioritize enumeration and are optimized to return one or a few best mapping locations.

Conversely, most prominent all-mappers are based on \emph{filtering algorithms} for approximate string matching \textcolor{black}{\cite{Navarro2001}}.
Seeds are sampled from given reads and used as anchors to quickly determine, with the help of an index, locations of the reference genome candidate to contain approximate occurrences.
Each candidate location \textcolor{black}{is} subsequently verified with an online method \textcolor{black}{\cite{Navarro2001b}}.
Increasing the error rate in filtering algorithms leads to a decrease of the seed length which in turn deteriorates filtration efficiency.
Current all-mappers \cite{Razers3, Ahmadi2011, Alkan2009, Shrimp2} are usually slower than best-mappers but conversely they are able to report all asked mapping locations in reasonable time.

\subsection{Our contribution}

\enlargethispage{-65.1pt}
\textcolor{black}
{
We present Masai, a read mapper that combines for the first time filtering with backtracking.
Our filtering approach is based on non-heuristic and full-sensitive filtration strategies using exact and \emph{approximate seeds}, which are searched in the reference genome via backtracking.
Approximate seeds, compared to exact seeds, increase filtration specificity while preserving sensitivity.
Moreover, we introduce a \emph{multiple backtracking} method which speeds up filtration by searching all seeds simultaneously with the help of an additional index.
Combined together, these methods yield a flexible and efficient filter that significantly speeds up approximate search on genomic datasets.}

\textcolor{black}
{
Masai targets all-mapping, but eventually it can be used as a best-mapper achieving even better runtimes.
We extensively compared Masai with popular read mappers on simulated and real datasets.
Compared to considered all-mappers, Masai is an order of magnitude faster and has comparable sensitivity.
In addition, Masai is more accurate than considered best-mappers and 2--3 times faster than Bowtie\,2 and \cite{Bowtie2} BWA \cite{BWA}.
Masai is implemented in \CC using the SeqAn library and distributed under the BSD license.
It can be downloaded from \url{http://www.seqan.de/projects/masai}.}


\section{Materials and Methods}

In order to map reads to a reference genome, we proceed as follows.

We first construct a conceptual \emph{suffix tree} of the reference genome, then store it on disk and reuse it for each read mapping job.
\textcolor{black}{
We choose the enhanced suffix array (Esa) \cite{Abouelhoda2004}, which provides an efficient implementation of the suffix tree and consumes 38 Gb of memory for the whole human genome.
However, any other data structure equivalent to the suffix tree in terms of allowing a prefix search, \ie the suffix array \cite{Manber1990} or the FM-index \cite{Ferragina2001} can be used to this intent.}

\begin{figure}[h]
\centering
\subcaptionbox{Exact seeds.}{\includegraphics[scale=0.9]{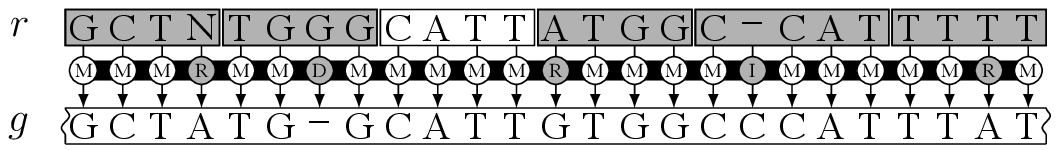}}
\\
\vspace{5mm}
\subcaptionbox{Approximate seeds.}{\includegraphics[scale=0.9]{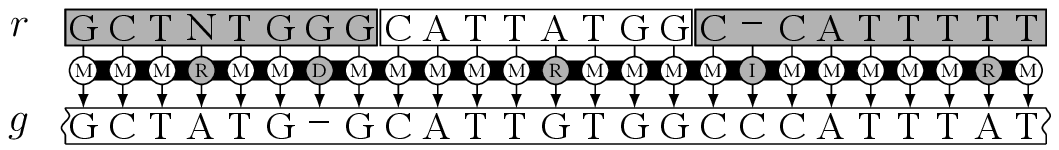}}
\caption{
{\bfseries Filtration strategies.}
A read $r$ occurs in the reference genome $g$ within edit distance 5.
(a) If we partition $r$ in 6 seeds, at least one seed (in white) occurs exactly in $g$.
(b) Alternatively, if we partition $r$ in 3 seeds, at least one seed (in white) occurs within edit distance 1 in $g$.
}
\label{fig:FiltrationStrategies}
\end{figure}

At mapping time we choose a filtration strategy according to the reference genome and the specified error rate.
Our filtration strategies are based on \cite{Navarro2000}, make use of \emph{exact \textcolor{black}{and} approximate non-overlapping seeds} and are guaranteed to be full-sensitive by the pigeonhole principle.
In Figure~\ref{fig:FiltrationStrategies} we show an example providing two alternative filtration strategies.

Therefore we partition all reads and their reverse complements in non-overlapping seeds and subsequently arrange all seeds in a conceptual \emph{radix tree}.
The time spent to construct the radix tree is easily justified since the tree allows us to perform multiple backtracking.
We indeed apply our \emph{multiple backtracking algorithm} to the radix tree, in order to search simultaneously all seeds in the suffix tree of the reference genome.

Finally we perform seed extension on each seed reported by the multiple backtracking algorithm.
We extend both ends of each seed using a banded version of \emph{Myers bit-vector algorithm} \cite{Myers1999} presented in \cite{Razers3}.

In the following of this section we give a detailed explanation of each mapping step.

\subsection{Seeds}
We now consider formally the read mapping problem.
Given a reference genome $g$, a set of reads $\mathcal{R}$ and an absolute number of errors $k$ consisting of indels and mismatches, for each read $r \in \mathcal{R}$ find all mapping locations where $r$ approximately occurs in $g$ within $k$ errors.

\subsubsection{Exact seeds}
A simple solution to the problem is provided by a filtering algorithm proposed in \cite{Baeza1999b} which reduces an approximate search into smaller exact searches.
Each read $r$ \textcolor{black}{is} partitioned into $k+1$ non-overlapping seeds which \textcolor{black}{are} searched in $g$ with the help of an index.
Since each edit operation can affect at most one seed, for the pigeonhole principle each approximate occurrence of $r$ in $g$ contains an exact occurrence of some seed.
However the converse is not true, consequently we must verify whether any candidate location induced by an occurrence of some seed corresponds to an approximate occurrence of $r$ in $g$.

Filtration specificity in terms of candidate locations to verify is strongly correlated to seed length.
Since we want to maximize the length of the shortest seed, we let the minimum seed length be $\lfloor |r|/(k+1) \rfloor$.
If we want to improve filtration specificity by increasing seed length, we resort to approximate seeds.

\subsubsection{Approximate seeds}
A more involved filtering algorithm proposed in \cite{Navarro2000} reduces an approximate search into smaller approximate searches.
We partition $r$ into $s \leq k+1$ non-overlapping seeds.
According to the pigeonhole principle each approximate occurrence of $r$ in $g$ then contains an approximate occurrence of some seed within distance $\lfloor k/s \rfloor$.

Moreover, we search $(k \bmod{s}) + 1$ seeds within distance $\lfloor k/s \rfloor$ and the remaining seeds within distance $\lfloor k/s \rfloor - 1$.
To prove full-sensitivity it suffices to see that, if none of the seeds occurs within its assigned distance, the total distance must be at least $s \cdot \lfloor k/s \rfloor + (k \bmod s) + 1 = k + 1$.
Hence all approximate occurrences of $r$ in $g$ within distance $k$ will be found.

Seeds \textcolor{black}{are} searched approximately by backtracking on a suffix tree.
We will introduce two efficient multiple backtracking algorithms to search exactly or approximately a set of seeds.

\subsubsection{Filtration strategies}
With approximate seeds we are free to choose the number of seeds $s$, which in turn enforces the minimum seed length $l$ to be $\lfloor |r|/s \rfloor$.
Or vice versa we fix $l$, which enforces $s$ to be $\lfloor |r|/l \rfloor$.
The resulting filter is flexible, indeed by increasing $l$ filtration becomes more specific at the expense of a higher filtration time.

The optimal seed length $l$ depends on the reference genome as well as on read length and the absolute number of errors.
\textcolor{black}{
When mapping current NGS datasets on short to medium length genomes, \eg bacterial genomes, exact seeds are still more efficient than approximate seeds.}
Conversely on larger genomes, \textcolor{black}{\eg} mammalian genomes, approximate seeds outperform exact seeds by an order of magnitude.
\textcolor{black}{Filtration results are provided in the Supplementary Data.}


\subsection{Indices}
\label{sub:Indices}

We make use of two fundamental data structures, radix and suffix trees.
Here we present these indices and give most important implementation details.

\subsubsection{Radix tree}

The radix tree \cite{Morrison1968} is a lexicographically ordered tree data structure representing a set of strings.
There is one node designated as the root and one leaf per string in the set.
Every internal node has more than one child and edges are labeled with non-empty strings.
Consequently, common prefixes are compressed and each path from the root to an internal node spells a different substring.

The radix tree for a set of strings can be built in time and space linear in the total length of the strings.
It is the ideal data structure to iterate a set of strings in lexicographical order and ask for the longest common prefix of any two strings.

\subsubsection{Suffix tree}

The suffix tree \cite{Weiner1973} of a string is the radix tree of all the suffixes of the string itself.
It can be built in time and space linear in the length of the string \cite{Ukkonen1995}.

The suffix tree \textcolor{black}{is} used for exact search.
A pattern is found by starting in the root node and following the path spelling the pattern.
If such path is found, each leaf below the last traversed node points to a distinct occurrence of the pattern in the text.

Approximate search \textcolor{black}{is} performed on the suffix tree by means of backtracking \cite{Ukkonen1993, Baeza1999}.
A preorder depth-first search on the suffix tree spells all substrings present in the text.
While visiting each branch of the suffix tree, the distance between the pattern and the text spelled along the path \textcolor{black}{is} incrementally computed.
If the pattern approximately matches the spelled text, each leaf below the last traversed node points to a distinct approximate occurrence of the pattern in the text.
Conversely, if the remaining suffix of the pattern can not lead to any approximate occurrence, the branch is pruned and the visit proceeds on the next branch.

\subsubsection{Implementation}

\textcolor{black}{
We replace the suffix tree with the \emph{enhanced suffix array} (Esa) \cite{Abouelhoda2004}, which preserves the asymptotic performances of the suffix tree and consumes, as implemented in SeqAn, $12n$ bytes for a sequence of length $n$. We construct the Esa in linear time using the algorithms proposed in \cite{Karkkainen2003,Kasai2001,Abouelhoda2004}.}
Therefore we use an Esa to index the reference genome.
Similarly to all read mappers relying on an index of the reference genome, we build the Esa of the reference genome only once, store it on disk, and reuse it for each mapping job.

\textcolor{black}{
We emulate the radix tree by means of the \emph{lazy suffix tree}.
We use the \emph{wotd}-algorithm \cite{Giegerich1999} in order to build a partial suffix tree only containing certain suffixes.
However, when performing multiple backtracking with exact seeds, the radix tree construction time dominates the overall filtration time.
Therefore in this case we resort to the \emph{$q$-gram index} to emulate the radix tree.
We build the $q$-gram index efficiently and in linear time by bucket sort.}
Below depth $q$ the properties of the radix tree are lost, however multiple backtracking is still applicable.

\subsection{Multiple backtracking}
\label{sub:MultipleBacktracking}

We now introduce a method for multiple off-line approximate string matching to search simultaneously a set of patterns in a text.
We start by introducing an algorithm for multiple off-line exact string matching and later extend it to approximate string matching.

For simplicity of exposition we describe the algorithms working on tries, although they \textcolor{black}{are easily extendable} to work on trees.
Hence in the following we assume the text sequence and the set of patterns to be preprocessed respectively using a suffix trie $G$ and a radix trie $S$.
Given a node $x$, we denote with $label(x)$ the label of the edge entering into $x$, and with $\mathcal{C}(x)$ and $\mathcal{L}(x)$ respectively the set of children and the set of leaves below $x$.

\subsubsection{Exact search}

Algorithm~\ref{MultipleExactSearch} takes as input two nodes $g$, $s$ respectively of $G$, $S$ and reports all pairs of leaves $(l_g, l_s) \in \mathcal{L}(g) \times \mathcal{L}(s)$ such that the path from $s$ to $l_s$ spells a prefix of the path from $g$ to $l_g$.

Consequently by applying Algorithm~\ref{MultipleExactSearch} on the root nodes of $G$, $S$ we obtain all pairs of leaves $(l_g, l_s)$ such that the pattern pointed by $l_s$ occurs in the text at the position pointed by $l_g$.

\vspace*{2mm}
\begin{algorithm}
\caption{Multiple exact search.}
\label{MultipleExactSearch}
\begin{algorithmic}[1]
\algnotext{EndFor}
\Procedure{Search}{$g,s$}
	\If {$s$ is a leaf}
		\State \Report $\mathcal{L}(g) \times s$
	\Else
		\ForAll {$c_s \in \mathcal{C}(s)$}
			\If {$\exists\,{c_g \in \mathcal{C}(g)}:\  label(c_g) = label(c_s)$}
				\State \Call{Search}{$c_g,c_s$}
			\EndIf
		\EndFor
	\EndIf
\EndProcedure
\end{algorithmic}
\end{algorithm}

\subsubsection{Approximate search}

Algorithm~\ref{MultipleApproximateSearch} takes an additional input argument $k$ which denotes the maximum number of mismatches left and computes the union of all paths within $k$ mismatches in the subtrees rooted in $g$, $s$. It reports all pairs of leaves $(l_g, l_s) \in \mathcal{L}(r) \times \mathcal{L}(s)$ such that the path from $s$ to $l_s$ spells a prefix of the path from $g$ to $l_g$ with at most $k$ mismatches.

Therefore by applying Algorithm~\ref{MultipleApproximateSearch} on the root nodes of $G$, $S$ we obtain all pairs of leaves $(l_g, l_s)$ such that the pattern pointed by $l_s$ occurs within $k$ mismatches in the text at the position pointed by $l_g$.

\vspace*{2mm}
\begin{algorithm}
\caption{Multiple approximate search.}
\label{MultipleApproximateSearch}
\begin{algorithmic}[1]
\algnotext{EndFor}
\Procedure{Search}{$g,s,k$}
	\If {$k = 0$}
		\State \Call{Search}{$g,s$}
	\Else
		\If {$s$ is a leaf}
			\State \Report $\mathcal{L}(g) \times s$
		\Else
			\ForAll {$c_g \in \mathcal{C}(g)$}
				\ForAll {$c_s \in \mathcal{C}(s)$}
					\If {$label(c_g) = label(c_s)$}
						\State \Call{Search}{$c_g,c_s,k$}
					\Else
						\State \Call{Search}{$c_g,c_s,k-1$}
					\EndIf
				\EndFor
			\EndFor
		\EndIf
	\EndIf
\EndProcedure
\end{algorithmic}
\end{algorithm}

\begin{figure}[h]
\centering
\vspace{5mm}
\subcaptionbox{Text.}{\includegraphics[scale=0.8]{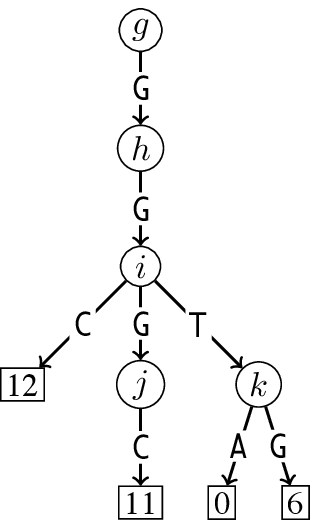}}
\hspace{10mm}
\subcaptionbox{Patterns.}{\includegraphics[scale=0.8]{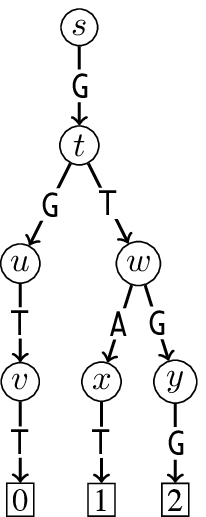}}
\hspace{10mm}
\subcaptionbox{States.}{\includegraphics[scale=0.8]{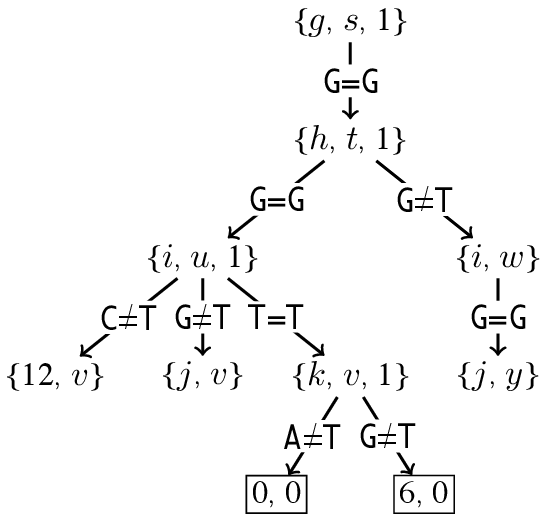}}
\caption{
{\bfseries Multiple backtracking.}
\textcolor{black}
{
(a) A part of the suffix trie representing the text GGTAACGGTGCGGGC (Supplementary Data). Numbers on the leaves are suffix positions in the text, while letters on the inner nodes are arbitrary and serve to distinguish nodes from each other.
(b) The trie representing the set of patterns \{GGTT, GTAT, GTGG\}, respectively numbered \{0, 1, 2\}. Labels on the leaves show pattern numbers, while labels on the inner nodes are again arbitrary identifiers.
(c) Recursive calls performed by Algorithm~\ref{MultipleApproximateSearch} called with arguments \{$g$, $s$, $1$\}. Edges represent comparisons performed by Algorithm~\ref{MultipleApproximateSearch} at line 10 or by Algorithm~\ref{MultipleExactSearch} at line 6, nodes with curly brackets represent recursive calls, rectangular leaves represent approximate matches reported.
In this example, pattern numbered 0 (GGTT) matches the text twice, at positions 0 and 6, within 1 mismatch.}
}

\label{fig:MSA}
\end{figure}

For $k=0$, lines 5--16 of Algorithm~\ref{MultipleApproximateSearch} are equivalent to Algorithm~\ref{MultipleExactSearch}.
However Algorithm~\ref{MultipleExactSearch} is preferred to Algorithm~\ref{MultipleApproximateSearch} because it traverses only edges spelling common strings instead of all pairs of edges and it is thus more efficient.
Figure~\ref{fig:MSA} depicts a run of Algorithm~\ref{MultipleApproximateSearch}.

Algorithm~\ref{MultipleApproximateSearch} only considers mismatches, but it can be extended to allow indels, \eg similarly to \cite{Navarro2000}.
In Masai Algorithm~\ref{MultipleApproximateSearch} is implemented only for mismatches, consequently full-sensitivity is not attained when using approximate seeds and considering mapping locations with indels.
However in the results section we show that such implementation detail sacrifices less than 1\% sensitivity.

\subsection{Seed extension}

We use a banded version of Myers bit-vector algorithm \cite{Myers1999} already presented in \cite{Razers3}.
\textcolor{black}{
Myers' algorithm is an efficient DP alignment algorithm \cite{Needleman1970} for edit distance. 
Instead of computing DP cells one after another, it encodes the whole DP column in two bit-vectors and computes the adjacent column in a constant number of 12 logical and 3 arithmetical operations.
We implemented a bit-parallel version that computes only a diagonal band of the DP matrix and is faster and more specific than the original algorithm by Myers.
More details can be found in the Supplementary Data.
}
However differently from \cite{Razers3}, instead of performing a semi-global alignment to verify a parallelogram surrounding the seed, we perform a global alignment on both ends of a seed.
Given a seed occurring with $e$ errors, we first perform seed extension on the left side within an error threshold of $k - e$ errors.
Only if the seed extension on the left side succeeds, we perform a seed extension on the right side within the remaining error threshold.
Moreover, we first compute the longest common prefix on each side of the seed and let the global alignment algorithm start from the first mismatching positions.
We observed that this approach is up to two times faster than \cite{Razers3}.


\section{Results}

We thoroughly compared Masai with the best-mappers Bowtie\,2, BWA and Soap\,2 as well as with the all-mappers RazerS\,3, Hobbes, mrFAST and SHRiMP\,2.
We remark that Bowtie\,2, BWA, Soap\,2 and SHRiMP\,2 rely on scoring schemes taking into account base quality values, while Masai, RazerS\,3, Hobbes and mrFAST use edit distance.
When relevant, read mappers that accept an absolute number of errors (Masai, mrFAST, Hobbes, Soap\,2) or an error rate (RazerS\,3) were configured accordingly.
We used default parameters, except where stated otherwise (Supplementary Data).

\textcolor{black}{We performed runtime experiments on real data.}
All read sets are given by their SRA/ENA id.
As references we used whole genomes of E.~coli (NCBI NC\_000913.2), C.~elegans (WormBase WS195), D.~melanogaster (FlyBase release 5.42), and H.~sapiens (GRCh37.p2).
The mapping times were measured on a cluster of nodes with 72\,GB RAM and 2 Intel Xeon X5650 processors running Linux~3.2.0.
For running time comparison, we ran the tools using a single thread and used local disks for I/O.

\subsection{Rabema benchmark}

We first used the Rabema benchmark~\cite{Holtgrewe2011} (v1.1) for a thorough evaluation and comparison of read mapping sensitivity.
\textcolor{black}{Similarly to \cite{Bowtie2}, }we used the read simulator Mason \cite{SeqAnReadSimulator} with default profile settings to simulate from each whole genome 100\,k reads of length 100\,bp having sequencing errors distributed like in a typical Illumina run.
Simulation details are included in Suppplementary Data.

The benchmark contains the categories \emph{all}, \emph{all-best}, \emph{any-best}, and \emph{recall}.
In the categories all, all-best, and any-best a read mapper had to find all, all of the best, or any of the best edit distance locations for each read.
The category recall required a read mapper to find the \emph{original} location of each read, which is a measure independent of the used scoring model, \eg edit distance or quality based.
We also classified mapping locations in each category by their edit distance.
The benchmark was performed for an error rate of 5\,\%, which corresponds to edit distance 5 for reads of length 100\,bp.

For a more fair and thorough comparison we also configured BWA and Bowtie\,2 as all-mappers (Soap\,2 could not be configured accordingly).
To this extent, we parametrized them to be highly sensitive and output all found mapping locations.
Since BWA and Bowtie\,2 were not designed to be used as all-mappers, they spent much more time than proper all-mappers, \ie up to 3~hours in a run compared to several minutes.
The aim here is to investigate read mapping sensitivity and therefore we do not report running times.

Results for H.~sapiens are shown in Table~\ref{tab:Rabema}.
Additional results for E.~coli, C.~elegans and D.~Melanogaster are shown in \textcolor{black}{the Supplementary Data}.

\begin{table*}[t]
  \caption[Rabema benchmark results]
  {
  \label{tab:Rabema}
    {\bfseries Rabema benchmark results.}
    Rabema scores in percent
    \textcolor{black}{(average fraction of edit distance locations reported per read).}
    Large numbers show total scores in each Rabema category and small numbers show the category scores separately for reads with $\bigl(\begin{smallmatrix}\mbox{\tiny 0}&\mbox{\tiny 1}&\mbox{\tiny 2}\\\mbox{\tiny 3}&\mbox{\tiny 4}&\mbox{\tiny 5}\end{smallmatrix}\bigr)$ errors.
    }
  \vspace{-3mm}
  \center
  \sffamily
  \resizebox{1.0\textwidth}{!}
  {
	\renewcommand{\tabcolsep}{0.8ex}
\begin{tabular}{llcccc}
  \toprule
   
  & method  &\multicolumn{1}{c}{  all } &\multicolumn{1}{c}{  all-best } &\multicolumn{1}{c}{  any-best } &\multicolumn{1}{c}{  recall } \\
    \midrule
	\multirow{4}{*}{\begin{sideways}\footnotesize best-mappers\hspace{1.1ex} \end{sideways}} &  Masai  & \cellcolor[rgb]{0.679089659394854,0.696527484523116,0.50698003842034}\phantom{0}93.26 {\subcolbeg\begin{tabular}{rrr} \cellcolor[rgb]{0.440479297869682,0.711440632118439,0.543434399208908}\phantom{0}99.18 & \cellcolor[rgb]{0.460193465894825,0.710208496616868,0.540422512427288}\phantom{0}98.73 & \cellcolor[rgb]{0.494378862359116,0.708071909337849,0.535199743523022}\phantom{0}97.93\\ \cellcolor[rgb]{0.590047872462076,0.702092596206414,0.520583644757292}\phantom{0}95.60 & \cellcolor[rgb]{0.922238285357935,0.681330695400423,0.46983233167598}\phantom{0}85.77 & \cellcolor[rgb]{0.954074606057408,0.496520811011724,0.484574365423691}\phantom{0}43.60\subcolvspace\\\end{tabular}\subcolend} & \cellcolor[rgb]{0.495280072867453,0.708015583681078,0.535062058584248}\phantom{0}97.91 {\subcolbeg\begin{tabular}{rrr} \cellcolor[rgb]{0.500704852054585,0.707676534981883,0.534233272875103}\phantom{0}97.79 & \cellcolor[rgb]{0.496507857724522,0.707938847127511,0.534874480342196}\phantom{0}97.88 & \cellcolor[rgb]{0.490375812684687,0.708322099942501,0.535811320556615}\phantom{0}98.03\\ \cellcolor[rgb]{0.492277055550075,0.708203272263414,0.535520852896625}\phantom{0}97.98 & \cellcolor[rgb]{0.483014574642471,0.70878217732014,0.536935954146398}\phantom{0}98.20 & \cellcolor[rgb]{0.545872389768145,0.704853563874785,0.527332676835531}\phantom{0}96.70\subcolvspace\\\end{tabular}\subcolend} & \cellcolor[rgb]{0.406062170820126,0.713591702559036,0.548692571397034}\phantom{0}99.95 {\subcolbeg\begin{tabular}{rrr} \cellcolor[rgb]{0.403921568627451,0.713725490196078,0.549019607843137}\phantom{}100.00 & \cellcolor[rgb]{0.403921568627451,0.713725490196078,0.549019607843137}\phantom{}100.00 & \cellcolor[rgb]{0.403921568627451,0.713725490196078,0.549019607843137}\phantom{}100.00\\ \cellcolor[rgb]{0.404774679166092,0.713672170787413,0.548889271510845}\phantom{0}99.98 & \cellcolor[rgb]{0.407024133280799,0.713531579905244,0.548545604909987}\phantom{0}99.93 & \cellcolor[rgb]{0.461183344773337,0.710146629186961,0.54027128093196}\phantom{0}98.71\subcolvspace\\\end{tabular}\subcolend} & \cellcolor[rgb]{0.502252928511953,0.707579780203297,0.533996761194116}\phantom{0}97.75 {\subcolbeg\begin{tabular}{rrr} \cellcolor[rgb]{0.496707942409465,0.707926341834703,0.534843911848663}\phantom{0}97.88 & \cellcolor[rgb]{0.498367204808318,0.707822637934774,0.534590413426616}\phantom{0}97.84 & \cellcolor[rgb]{0.500534371076229,0.70768719004303,0.534259318580129}\phantom{0}97.79\\ \cellcolor[rgb]{0.505110058042338,0.707401209607648,0.533560255293641}\phantom{0}97.68 & \cellcolor[rgb]{0.510073875080216,0.707090971042781,0.532801894357298}\phantom{0}97.56 & \cellcolor[rgb]{0.544229851520043,0.704956222515291,0.527583620178991}\phantom{0}96.74\subcolvspace\\\end{tabular}\subcolend} \\ 
    &  Bowtie\,2  & \cellcolor[rgb]{0.722804530220145,0.693795305096535,0.500301377599809}\phantom{0}92.04 {\subcolbeg\begin{tabular}{rrr} \cellcolor[rgb]{0.440479297869682,0.711440632118439,0.543434399208908}\phantom{0}99.18 & \cellcolor[rgb]{0.460500660775921,0.710189296936799,0.54037557987601}\phantom{0}98.72 & \cellcolor[rgb]{0.541772229620815,0.705109823883993,0.527959090191373}\phantom{0}96.80\\ \cellcolor[rgb]{0.672459153899506,0.696941891116575,0.507993032315462}\phantom{0}93.44 & \cellcolor[rgb]{0.967081073276227,0.659101651246955,0.465064664595463}\phantom{0}81.94 & \cellcolor[rgb]{0.953759536552322,0.492582442198139,0.485046969681321}\phantom{0}40.19\subcolvspace\\\end{tabular}\subcolend} & \cellcolor[rgb]{0.567610083970622,0.70349495798713,0.524011640221264}\phantom{0}96.16 {\subcolbeg\begin{tabular}{rrr} \cellcolor[rgb]{0.500704852054585,0.707676534981883,0.534233272875103}\phantom{0}97.79 & \cellcolor[rgb]{0.497841873093291,0.707855471166963,0.534670672438634}\phantom{0}97.85 & \cellcolor[rgb]{0.582164502653447,0.702585306819454,0.521788048478054}\phantom{0}95.80\\ \cellcolor[rgb]{0.619892916670176,0.700227280943408,0.516023985225499}\phantom{0}94.83 & \cellcolor[rgb]{0.674893510555194,0.696789743825595,0.507621116715288}\phantom{0}93.37 & \cellcolor[rgb]{0.829043491855854,0.687155369994303,0.484070425127687}\phantom{0}88.86\subcolvspace\\\end{tabular}\subcolend} & \cellcolor[rgb]{0.488232868538503,0.708456033951638,0.536138714801171}\phantom{0}98.08 {\subcolbeg\begin{tabular}{rrr} \cellcolor[rgb]{0.403921568627451,0.713725490196078,0.549019607843137}\phantom{}100.00 & \cellcolor[rgb]{0.405530531638357,0.713624930007897,0.548773794049804}\phantom{0}99.96 & \cellcolor[rgb]{0.510552970381555,0.707061027586447,0.532728699241816}\phantom{0}97.55\\ \cellcolor[rgb]{0.549097637796711,0.704651985873,0.526839930608945}\phantom{0}96.62 & \cellcolor[rgb]{0.616266287051791,0.700453945294557,0.516578053639419}\phantom{0}94.93 & \cellcolor[rgb]{0.776908163283075,0.690413828030102,0.49203554477075}\phantom{0}90.46\subcolvspace\\\end{tabular}\subcolend} & \cellcolor[rgb]{0.576488511190601,0.702940056285882,0.522655213840434}\phantom{0}95.94 {\subcolbeg\begin{tabular}{rrr} \cellcolor[rgb]{0.490992978940581,0.708283527051508,0.53571703126752}\phantom{0}98.01 & \cellcolor[rgb]{0.50340046541151,0.707508059147075,0.533821443056684}\phantom{0}97.72 & \cellcolor[rgb]{0.591886329692351,0.701977692629522,0.520302769347111}\phantom{0}95.55\\ \cellcolor[rgb]{0.642645022648344,0.698805274319773,0.51254796903439}\phantom{0}94.24 & \cellcolor[rgb]{0.695922656692039,0.695475422192042,0.504408330499936}\phantom{0}92.79 & \cellcolor[rgb]{0.807935791824488,0.688474601246264,0.487295212632479}\phantom{0}89.52\subcolvspace\\\end{tabular}\subcolend} \\ 
     &  BWA  & \cellcolor[rgb]{0.71779743402355,0.694108248608822,0.501066350629844}\phantom{0}92.18 {\subcolbeg\begin{tabular}{rrr} \cellcolor[rgb]{0.440479297869682,0.711440632118439,0.543434399208908}\phantom{0}99.18 & \cellcolor[rgb]{0.460525565578098,0.710187740386663,0.540371774975677}\phantom{0}98.72 & \cellcolor[rgb]{0.499521574896626,0.707750489804255,0.534414051329791}\phantom{0}97.81\\ \cellcolor[rgb]{0.642182510812693,0.698834181309501,0.512618630564836}\phantom{0}94.25 & \cellcolor[rgb]{0.966392879716893,0.650499231755276,0.466096954934465}\phantom{0}80.92 & \cellcolor[rgb]{0.953571551984206,0.490232635096695,0.485328946533495}\phantom{0}37.65\subcolvspace\\\end{tabular}\subcolend} & \cellcolor[rgb]{0.541386731157968,0.705133917537921,0.528017985789864}\phantom{0}96.81 {\subcolbeg\begin{tabular}{rrr} \cellcolor[rgb]{0.500704852054585,0.707676534981883,0.534233272875103}\phantom{0}97.79 & \cellcolor[rgb]{0.497245950758136,0.707892716312911,0.534761716128727}\phantom{0}97.87 & \cellcolor[rgb]{0.496815929971905,0.70791959261205,0.534827413748846}\phantom{0}97.88\\ \cellcolor[rgb]{0.550259348779752,0.70457937893656,0.526662446986536}\phantom{0}96.59 & \cellcolor[rgb]{0.701814898280047,0.695107157092791,0.503508126923991}\phantom{0}92.63 & \cellcolor[rgb]{0.968167984441338,0.672688040810846,0.463434297847797}\phantom{0}83.47\subcolvspace\\\end{tabular}\subcolend} & \cellcolor[rgb]{0.456713667088015,0.710425984042293,0.540954148356107}\phantom{0}98.81 {\subcolbeg\begin{tabular}{rrr} \cellcolor[rgb]{0.403921568627451,0.713725490196078,0.549019607843137}\phantom{}100.00 & \cellcolor[rgb]{0.406173635254615,0.713584736031881,0.548675542108432}\phantom{0}99.95 & \cellcolor[rgb]{0.412447526860781,0.713192617806495,0.547717030890823}\phantom{0}99.81\\ \cellcolor[rgb]{0.468209981343603,0.709707464401319,0.539197767011503}\phantom{0}98.55 & \cellcolor[rgb]{0.64098471631688,0.698909043465489,0.512801626946141}\phantom{0}94.28 & \cellcolor[rgb]{0.933581310365173,0.680621756337471,0.468099369522096}\phantom{0}85.37\subcolvspace\\\end{tabular}\subcolend} & \cellcolor[rgb]{0.557388706437161,0.704133794082972,0.525573239566654}\phantom{0}96.41 {\subcolbeg\begin{tabular}{rrr} \cellcolor[rgb]{0.494710385371942,0.708051189149548,0.53514909417384}\phantom{0}97.93 & \cellcolor[rgb]{0.504617880016039,0.707431970734292,0.53363544915877}\phantom{0}97.69 & \cellcolor[rgb]{0.52303974570493,0.706280604128736,0.5308209974563}\phantom{0}97.25\\ \cellcolor[rgb]{0.583233025218664,0.702518524159128,0.521624801975035}\phantom{0}95.77 & \cellcolor[rgb]{0.724867729493042,0.693666355141979,0.499986166599783}\phantom{0}91.98 & \cellcolor[rgb]{0.954441259496993,0.679318009516732,0.464912432849179}\phantom{0}84.61\subcolvspace\\\end{tabular}\subcolend} \\ 
      &  Soap\,2  & \cellcolor[rgb]{0.958868452356375,0.556443889748805,0.477383595975241}\phantom{0}65.93 {\subcolbeg\begin{tabular}{rrr} \cellcolor[rgb]{0.440479297869682,0.711440632118439,0.543434399208908}\phantom{0}99.18 & \cellcolor[rgb]{0.591942229494589,0.701974198891882,0.520294229099547}\phantom{0}95.55 & \cellcolor[rgb]{0.747229524772314,0.692268742937025,0.496569781209894}\phantom{0}91.34\\ \cellcolor[rgb]{0.95294295275667,0.482375144752487,0.4862718453748}\phantom{00}8.67 & \cellcolor[rgb]{0.952941176544256,0.482352942097313,0.48627450969342}\phantom{00}0.70 & \cellcolor[rgb]{0.952941176470588,0.482352941176471,0.486274509803922}\phantom{00}0.00\subcolvspace\\\end{tabular}\subcolend} & \cellcolor[rgb]{0.960426039612903,0.57591373045541,0.475047215090449}\phantom{0}69.89 {\subcolbeg\begin{tabular}{rrr} \cellcolor[rgb]{0.500704852054585,0.707676534981883,0.534233272875103}\phantom{0}97.79 & \cellcolor[rgb]{0.623584746314713,0.699996541590625,0.515459955696472}\phantom{0}94.74 & \cellcolor[rgb]{0.74611495155065,0.692338403763379,0.496740063229871}\phantom{0}91.37\\ \cellcolor[rgb]{0.952943216853462,0.48237844596239,0.486271449229611}\phantom{00}8.98 & \cellcolor[rgb]{0.952941176590615,0.48235294267681,0.486274509623881}\phantom{00}0.79 & \cellcolor[rgb]{0.952941176470588,0.482352941176471,0.486274509803922}\phantom{00}0.00\subcolvspace\\\end{tabular}\subcolend} & \cellcolor[rgb]{0.961082747735752,0.584122581991022,0.474062152906175}\phantom{0}71.37 {\subcolbeg\begin{tabular}{rrr} \cellcolor[rgb]{0.403921568627451,0.713725490196078,0.549019607843137}\phantom{}100.00 & \cellcolor[rgb]{0.54271341013705,0.705051000101728,0.527815298723615}\phantom{0}96.78 & \cellcolor[rgb]{0.681957369832253,0.696348252620778,0.506541915992404}\phantom{0}93.18\\ \cellcolor[rgb]{0.952943435419163,0.482381178033659,0.486271121381059}\phantom{00}9.21 & \cellcolor[rgb]{0.952941176606452,0.482352942874767,0.486274509600126}\phantom{00}0.81 & \cellcolor[rgb]{0.952941176470588,0.482352941176471,0.486274509803922}\phantom{00}0.00\subcolvspace\\\end{tabular}\subcolend} & \cellcolor[rgb]{0.96043633958237,0.576042480073743,0.475031765136249}\phantom{0}69.91 {\subcolbeg\begin{tabular}{rrr} \cellcolor[rgb]{0.489273891310443,0.708390970028391,0.535979669655458}\phantom{0}98.05 & \cellcolor[rgb]{0.628220101046454,0.699706831919891,0.514751776501345}\phantom{0}94.62 & \cellcolor[rgb]{0.751888923657381,0.691977530506708,0.495857928602454}\phantom{0}91.20\\ \cellcolor[rgb]{0.952947371392574,0.482430377701295,0.486265217420943}\phantom{0}11.85 & \cellcolor[rgb]{0.952941177717042,0.482352956757144,0.486274507934241}\phantom{00}1.41 & \cellcolor[rgb]{0.952941176476002,0.482352941244141,0.486274509795801}\phantom{00}0.36\subcolvspace\\\end{tabular}\subcolend} \\ 
		\midrule\multirow{7}{*}{\begin{sideways}\footnotesize all-mappers\hspace{3ex} \end{sideways}} &  Masai   & \cellcolor[rgb]{0.408448912779388,0.713442531186582,0.548327930264369}\phantom{0}99.90 {\subcolbeg\begin{tabular}{rrr} \cellcolor[rgb]{0.403921568627451,0.713725490196078,0.549019607843137}\phantom{}100.00 & \cellcolor[rgb]{0.403921568627451,0.713725490196078,0.549019607843137}\phantom{}100.00 & \cellcolor[rgb]{0.403921568627451,0.713725490196078,0.549019607843137}\phantom{}100.00\\ \cellcolor[rgb]{0.403933010669245,0.713724775068466,0.549017859753419}\phantom{}100.00 & \cellcolor[rgb]{0.406613761210927,0.713557228159611,0.548608300642884}\phantom{0}99.94 & \cellcolor[rgb]{0.466817282861163,0.709794508056471,0.539410540390765}\phantom{0}98.58\subcolvspace\\\end{tabular}\subcolend} & \cellcolor[rgb]{0.405919558750637,0.713600615813379,0.548714359352095}\phantom{0}99.96 {\subcolbeg\begin{tabular}{rrr} \cellcolor[rgb]{0.403921568627451,0.713725490196078,0.549019607843137}\phantom{}100.00 & \cellcolor[rgb]{0.403921568627451,0.713725490196078,0.549019607843137}\phantom{}100.00 & \cellcolor[rgb]{0.403921568627451,0.713725490196078,0.549019607843137}\phantom{}100.00\\ \cellcolor[rgb]{0.403921568627451,0.713725490196078,0.549019607843137}\phantom{}100.00 & \cellcolor[rgb]{0.407024133280799,0.713531579905244,0.548545604909987}\phantom{0}99.93 & \cellcolor[rgb]{0.461183344773337,0.710146629186961,0.54027128093196}\phantom{0}98.71\subcolvspace\\\end{tabular}\subcolend} & \cellcolor[rgb]{0.405919558750637,0.713600615813379,0.548714359352095}\phantom{0}99.96 {\subcolbeg\begin{tabular}{rrr} \cellcolor[rgb]{0.403921568627451,0.713725490196078,0.549019607843137}\phantom{}100.00 & \cellcolor[rgb]{0.403921568627451,0.713725490196078,0.549019607843137}\phantom{}100.00 & \cellcolor[rgb]{0.403921568627451,0.713725490196078,0.549019607843137}\phantom{}100.00\\ \cellcolor[rgb]{0.403921568627451,0.713725490196078,0.549019607843137}\phantom{}100.00 & \cellcolor[rgb]{0.407024133280799,0.713531579905244,0.548545604909987}\phantom{0}99.93 & \cellcolor[rgb]{0.461183344773337,0.710146629186961,0.54027128093196}\phantom{0}98.71\subcolvspace\\\end{tabular}\subcolend} & \cellcolor[rgb]{0.405924007863707,0.713600337743812,0.548713679626487}\phantom{0}99.96 {\subcolbeg\begin{tabular}{rrr} \cellcolor[rgb]{0.403921568627451,0.713725490196078,0.549019607843137}\phantom{}100.00 & \cellcolor[rgb]{0.403921568627451,0.713725490196078,0.549019607843137}\phantom{}100.00 & \cellcolor[rgb]{0.403921568627451,0.713725490196078,0.549019607843137}\phantom{}100.00\\ \cellcolor[rgb]{0.403921568627451,0.713725490196078,0.549019607843137}\phantom{}100.00 & \cellcolor[rgb]{0.406927120191574,0.713537643223321,0.548560426354174}\phantom{0}99.93 & \cellcolor[rgb]{0.457997693817885,0.710345732371676,0.54075797760571}\phantom{0}98.78\subcolvspace\\\end{tabular}\subcolend} \\ 
       &  Bowtie\,2   & \cellcolor[rgb]{0.586504981130466,0.70231402691464,0.521124919821843}\phantom{0}95.69 {\subcolbeg\begin{tabular}{rrr} \cellcolor[rgb]{0.405037352103444,0.713655753728829,0.548849140923194}\phantom{0}99.98 & \cellcolor[rgb]{0.40792992889796,0.713474967679172,0.548407219468476}\phantom{0}99.91 & \cellcolor[rgb]{0.428591959938626,0.71218359073913,0.545250520281708}\phantom{0}99.45\\ \cellcolor[rgb]{0.49210276642262,0.70821416533388,0.535547480402209}\phantom{0}97.99 & \cellcolor[rgb]{0.769381978365176,0.690884214587471,0.493185378577652}\phantom{0}90.69 & \cellcolor[rgb]{0.955841123429237,0.518602278159579,0.481924589365949}\phantom{0}55.14\subcolvspace\\\end{tabular}\subcolend} & \cellcolor[rgb]{0.455159216706129,0.710523137191161,0.541191633831117}\phantom{0}98.85 {\subcolbeg\begin{tabular}{rrr} \cellcolor[rgb]{0.415746837112243,0.712986410915779,0.547212969602405}\phantom{0}99.74 & \cellcolor[rgb]{0.413260034380837,0.713141836086492,0.547592897797481}\phantom{0}99.79 & \cellcolor[rgb]{0.465290663221998,0.709889921783919,0.539643773946748}\phantom{0}98.61\\ \cellcolor[rgb]{0.482621033151985,0.708806773663295,0.536996078540778}\phantom{0}98.21 & \cellcolor[rgb]{0.510603240164727,0.707057885724999,0.532721019136054}\phantom{0}97.55 & \cellcolor[rgb]{0.657377092501617,0.697884519953943,0.51029723614014}\phantom{0}93.84\subcolvspace\\\end{tabular}\subcolend} & \cellcolor[rgb]{0.441385732403407,0.711383979960081,0.543295916155144}\phantom{0}99.16 {\subcolbeg\begin{tabular}{rrr} \cellcolor[rgb]{0.403921568627451,0.713725490196078,0.549019607843137}\phantom{}100.00 & \cellcolor[rgb]{0.404887152868133,0.713665141181036,0.548872088028589}\phantom{0}99.98 & \cellcolor[rgb]{0.447933812529625,0.710974724952193,0.54229551502475}\phantom{0}99.01\\ \cellcolor[rgb]{0.464666727881163,0.709928917742721,0.539739097401598}\phantom{0}98.63 & \cellcolor[rgb]{0.494251541950389,0.708079866863395,0.535219195252133}\phantom{0}97.94 & \cellcolor[rgb]{0.645264831522307,0.69864153626515,0.512147720456423}\phantom{0}94.17\subcolvspace\\\end{tabular}\subcolend} & \cellcolor[rgb]{0.468243167408687,0.709705390272251,0.539192696918226}\phantom{0}98.54 {\subcolbeg\begin{tabular}{rrr} \cellcolor[rgb]{0.415432947143959,0.713006029038797,0.547260925014226}\phantom{0}99.74 & \cellcolor[rgb]{0.422566687068881,0.712560170293489,0.546171048081252}\phantom{0}99.58 & \cellcolor[rgb]{0.479993324880718,0.708971005430249,0.53739753397111}\phantom{0}98.27\\ \cellcolor[rgb]{0.506961342124918,0.707285504352487,0.533277420225469}\phantom{0}97.64 & \cellcolor[rgb]{0.538822308784052,0.705294193936291,0.528409772541434}\phantom{0}96.87 & \cellcolor[rgb]{0.63649114254394,0.699189891826298,0.513488145161451}\phantom{0}94.40\subcolvspace\\\end{tabular}\subcolend} \\ 
        &  BWA   & \cellcolor[rgb]{0.578289446830269,0.702827497808402,0.522380070895485}\phantom{0}95.89 {\subcolbeg\begin{tabular}{rrr} \cellcolor[rgb]{0.405831471160767,0.713606121287746,0.548727817178325}\phantom{0}99.96 & \cellcolor[rgb]{0.409235590210818,0.713393363847118,0.548207743434567}\phantom{0}99.88 & \cellcolor[rgb]{0.426762531532277,0.712297930014527,0.545530016288233}\phantom{0}99.49\\ \cellcolor[rgb]{0.528177931053981,0.70595946754442,0.530035996916862}\phantom{0}97.13 & \cellcolor[rgb]{0.8625288681358,0.685062533976807,0.478954603751584}\phantom{0}87.79 & \cellcolor[rgb]{0.958241405562947,0.548605804830955,0.478324166165383}\phantom{0}64.11\subcolvspace\\\end{tabular}\subcolend} & \cellcolor[rgb]{0.492471034068564,0.708191148606009,0.535491217289634}\phantom{0}97.98 {\subcolbeg\begin{tabular}{rrr} \cellcolor[rgb]{0.456580445399722,0.710434310397812,0.540974501669596}\phantom{0}98.81 & \cellcolor[rgb]{0.447848402010298,0.710980063109651,0.542308563854091}\phantom{0}99.01 & \cellcolor[rgb]{0.447696676878587,0.710989545930382,0.542331744082547}\phantom{0}99.02\\ \cellcolor[rgb]{0.498961410506439,0.707785500078642,0.534499632000514}\phantom{0}97.83 & \cellcolor[rgb]{0.65327848593817,0.698140682864159,0.510923412142889}\phantom{0}93.95 & \cellcolor[rgb]{0.938318116071603,0.680325705980819,0.467375690872503}\phantom{0}85.20\subcolvspace\\\end{tabular}\subcolend} & \cellcolor[rgb]{0.45630029143901,0.710451820020356,0.541017302969149}\phantom{0}98.82 {\subcolbeg\begin{tabular}{rrr} \cellcolor[rgb]{0.403921568627451,0.713725490196078,0.549019607843137}\phantom{}100.00 & \cellcolor[rgb]{0.406012885141944,0.713594782913923,0.548700101153423}\phantom{0}99.95 & \cellcolor[rgb]{0.412100476532403,0.713214308452019,0.54777005246877}\phantom{0}99.82\\ \cellcolor[rgb]{0.467665439763723,0.709741498250061,0.539280960863985}\phantom{0}98.56 & \cellcolor[rgb]{0.638900703501885,0.699039294266426,0.513120017792877}\phantom{0}94.34 & \cellcolor[rgb]{0.933581310365173,0.680621756337471,0.468099369522096}\phantom{0}85.37\subcolvspace\\\end{tabular}\subcolend} & \cellcolor[rgb]{0.499978573381396,0.707721927398957,0.53434423211684}\phantom{0}97.80 {\subcolbeg\begin{tabular}{rrr} \cellcolor[rgb]{0.447078781516443,0.711028164390516,0.542426144762875}\phantom{0}99.03 & \cellcolor[rgb]{0.450021257758656,0.710844259625378,0.541976599781426}\phantom{0}98.96 & \cellcolor[rgb]{0.45948604983543,0.71025271012058,0.540530589880807}\phantom{0}98.75\\ \cellcolor[rgb]{0.518802833395791,0.706545411148057,0.53146830350353}\phantom{0}97.35 & \cellcolor[rgb]{0.672605289871955,0.696932757618297,0.507970705986338}\phantom{0}93.43 & \cellcolor[rgb]{0.905151157517741,0.682398640890435,0.47244286509601}\phantom{0}86.36\subcolvspace\\\end{tabular}\subcolend} \\ 
         &  RazerS\,3   & \cellcolor[rgb]{0.403921568627451,0.713725490196078,0.549019607843137}\phantom{}100.00 {\subcolbeg\begin{tabular}{rrr} \cellcolor[rgb]{0.403921568627451,0.713725490196078,0.549019607843137}\phantom{}100.00 & \cellcolor[rgb]{0.403921568627451,0.713725490196078,0.549019607843137}\phantom{}100.00 & \cellcolor[rgb]{0.403921568627451,0.713725490196078,0.549019607843137}\phantom{}100.00\\ \cellcolor[rgb]{0.403921568627451,0.713725490196078,0.549019607843137}\phantom{}100.00 & \cellcolor[rgb]{0.403921568627451,0.713725490196078,0.549019607843137}\phantom{}100.00 & \cellcolor[rgb]{0.403921568627451,0.713725490196078,0.549019607843137}\phantom{}100.00\subcolvspace\\\end{tabular}\subcolend} & \cellcolor[rgb]{0.403921568627451,0.713725490196078,0.549019607843137}\phantom{}100.00 {\subcolbeg\begin{tabular}{rrr} \cellcolor[rgb]{0.403921568627451,0.713725490196078,0.549019607843137}\phantom{}100.00 & \cellcolor[rgb]{0.403921568627451,0.713725490196078,0.549019607843137}\phantom{}100.00 & \cellcolor[rgb]{0.403921568627451,0.713725490196078,0.549019607843137}\phantom{}100.00\\ \cellcolor[rgb]{0.403921568627451,0.713725490196078,0.549019607843137}\phantom{}100.00 & \cellcolor[rgb]{0.403921568627451,0.713725490196078,0.549019607843137}\phantom{}100.00 & \cellcolor[rgb]{0.403921568627451,0.713725490196078,0.549019607843137}\phantom{}100.00\subcolvspace\\\end{tabular}\subcolend} & \cellcolor[rgb]{0.403921568627451,0.713725490196078,0.549019607843137}\phantom{}100.00 {\subcolbeg\begin{tabular}{rrr} \cellcolor[rgb]{0.403921568627451,0.713725490196078,0.549019607843137}\phantom{}100.00 & \cellcolor[rgb]{0.403921568627451,0.713725490196078,0.549019607843137}\phantom{}100.00 & \cellcolor[rgb]{0.403921568627451,0.713725490196078,0.549019607843137}\phantom{}100.00\\ \cellcolor[rgb]{0.403921568627451,0.713725490196078,0.549019607843137}\phantom{}100.00 & \cellcolor[rgb]{0.403921568627451,0.713725490196078,0.549019607843137}\phantom{}100.00 & \cellcolor[rgb]{0.403921568627451,0.713725490196078,0.549019607843137}\phantom{}100.00\subcolvspace\\\end{tabular}\subcolend} & \cellcolor[rgb]{0.403921568627451,0.713725490196078,0.549019607843137}\phantom{}100.00 {\subcolbeg\begin{tabular}{rrr} \cellcolor[rgb]{0.403921568627451,0.713725490196078,0.549019607843137}\phantom{}100.00 & \cellcolor[rgb]{0.403921568627451,0.713725490196078,0.549019607843137}\phantom{}100.00 & \cellcolor[rgb]{0.403921568627451,0.713725490196078,0.549019607843137}\phantom{}100.00\\ \cellcolor[rgb]{0.403921568627451,0.713725490196078,0.549019607843137}\phantom{}100.00 & \cellcolor[rgb]{0.403921568627451,0.713725490196078,0.549019607843137}\phantom{}100.00 & \cellcolor[rgb]{0.403921568627451,0.713725490196078,0.549019607843137}\phantom{}100.00\subcolvspace\\\end{tabular}\subcolend} \\ 
          &  Hobbes   & \cellcolor[rgb]{0.55144296358397,0.704505403011296,0.526481616947002}\phantom{0}96.56 {\subcolbeg\begin{tabular}{rrr} \cellcolor[rgb]{0.430183364041121,0.712084127982724,0.545007389099382}\phantom{0}99.41 & \cellcolor[rgb]{0.448255094590534,0.710954644823386,0.542246430265444}\phantom{0}99.00 & \cellcolor[rgb]{0.458789068106892,0.710296271478613,0.540637073200445}\phantom{0}98.76\\ \cellcolor[rgb]{0.500201455562573,0.707707997262633,0.534310180672494}\phantom{0}97.80 & \cellcolor[rgb]{0.681006040056348,0.696407710731772,0.5066872580415}\phantom{0}93.20 & \cellcolor[rgb]{0.961873358322616,0.594005214326817,0.47287623702588}\phantom{0}73.05\subcolvspace\\\end{tabular}\subcolend} & \cellcolor[rgb]{0.529998250160197,0.705845697600282,0.529757892608968}\phantom{0}97.08 {\subcolbeg\begin{tabular}{rrr} \cellcolor[rgb]{0.524021188606496,0.706219263947388,0.530671054790783}\phantom{0}97.23 & \cellcolor[rgb]{0.550254029614557,0.704579711384384,0.526663259636774}\phantom{0}96.59 & \cellcolor[rgb]{0.532859942397727,0.705666841835436,0.529320689628234}\phantom{0}97.01\\ \cellcolor[rgb]{0.517650374739979,0.706617439814045,0.531644373575946}\phantom{0}97.38 & \cellcolor[rgb]{0.484925031513235,0.708662773765717,0.536644078791142}\phantom{0}98.16 & \cellcolor[rgb]{0.515860293785056,0.706729319873728,0.531917858166281}\phantom{0}97.42\subcolvspace\\\end{tabular}\subcolend} & \cellcolor[rgb]{0.49137315853606,0.70825976582679,0.535658948273766}\phantom{0}98.01 {\subcolbeg\begin{tabular}{rrr} \cellcolor[rgb]{0.494830942788078,0.708043654311039,0.535130675679708}\phantom{0}97.92 & \cellcolor[rgb]{0.512164074959891,0.706960333550301,0.53248255826457}\phantom{0}97.51 & \cellcolor[rgb]{0.493109997248757,0.708151213407247,0.535393597914882}\phantom{0}97.96\\ \cellcolor[rgb]{0.473372724148446,0.709384792976016,0.538409014638541}\phantom{0}98.43 & \cellcolor[rgb]{0.443153712982287,0.711273481173901,0.543025808011148}\phantom{0}99.12 & \cellcolor[rgb]{0.471766799739819,0.709485163251555,0.53865436420097}\phantom{0}98.46\subcolvspace\\\end{tabular}\subcolend} & \cellcolor[rgb]{0.557730734361691,0.704112417337688,0.525520985300406}\phantom{0}96.41 {\subcolbeg\begin{tabular}{rrr} \cellcolor[rgb]{0.59429230783762,0.701827318995443,0.519935189352695}\phantom{0}95.49 & \cellcolor[rgb]{0.580458259083513,0.702691947042575,0.522048724579017}\phantom{0}95.84 & \cellcolor[rgb]{0.552314133798678,0.704450954872877,0.526348521497533}\phantom{0}96.54\\ \cellcolor[rgb]{0.532358392009158,0.705698188734722,0.529397315382043}\phantom{0}97.03 & \cellcolor[rgb]{0.492641530492104,0.708180492579538,0.535465169224926}\phantom{0}97.98 & \cellcolor[rgb]{0.50039805080678,0.70769571005987,0.534280145287962}\phantom{0}97.79\subcolvspace\\\end{tabular}\subcolend} \\ 
           &  mrFAST   & \cellcolor[rgb]{0.405418407885201,0.713631937742469,0.548790924067648}\phantom{0}99.97 {\subcolbeg\begin{tabular}{rrr} \cellcolor[rgb]{0.403921568627451,0.713725490196078,0.549019607843137}\phantom{}100.00 & \cellcolor[rgb]{0.403921568627451,0.713725490196078,0.549019607843137}\phantom{}100.00 & \cellcolor[rgb]{0.403921568627451,0.713725490196078,0.549019607843137}\phantom{}100.00\\ \cellcolor[rgb]{0.404067448148125,0.713716372726036,0.548997320694145}\phantom{}100.00 & \cellcolor[rgb]{0.404511069697887,0.713688646379176,0.548929545179598}\phantom{0}99.99 & \cellcolor[rgb]{0.425028615645792,0.712406299757432,0.545794920104224}\phantom{0}99.53\subcolvspace\\\end{tabular}\subcolend} & \cellcolor[rgb]{0.405169196472516,0.713647513455762,0.548828998033475}\phantom{0}99.97 {\subcolbeg\begin{tabular}{rrr} \cellcolor[rgb]{0.403921568627451,0.713725490196078,0.549019607843137}\phantom{}100.00 & \cellcolor[rgb]{0.403921568627451,0.713725490196078,0.549019607843137}\phantom{}100.00 & \cellcolor[rgb]{0.403921568627451,0.713725490196078,0.549019607843137}\phantom{}100.00\\ \cellcolor[rgb]{0.404046721763259,0.71371766812509,0.549000487225167}\phantom{}100.00 & \cellcolor[rgb]{0.403921568627451,0.713725490196078,0.549019607843137}\phantom{}100.00 & \cellcolor[rgb]{0.444054748210841,0.711217166472117,0.542888149851231}\phantom{0}99.10\subcolvspace\\\end{tabular}\subcolend} & \cellcolor[rgb]{0.405111168078438,0.713651140230392,0.54883786348257}\phantom{0}99.97 {\subcolbeg\begin{tabular}{rrr} \cellcolor[rgb]{0.403921568627451,0.713725490196078,0.549019607843137}\phantom{}100.00 & \cellcolor[rgb]{0.403921568627451,0.713725490196078,0.549019607843137}\phantom{}100.00 & \cellcolor[rgb]{0.403921568627451,0.713725490196078,0.549019607843137}\phantom{}100.00\\ \cellcolor[rgb]{0.403921568627451,0.713725490196078,0.549019607843137}\phantom{}100.00 & \cellcolor[rgb]{0.403921568627451,0.713725490196078,0.549019607843137}\phantom{}100.00 & \cellcolor[rgb]{0.442856409371815,0.711292062649556,0.543071229396082}\phantom{0}99.13\subcolvspace\\\end{tabular}\subcolend} & \cellcolor[rgb]{0.405161488109598,0.713647995228444,0.548830175700031}\phantom{0}99.97 {\subcolbeg\begin{tabular}{rrr} \cellcolor[rgb]{0.403921568627451,0.713725490196078,0.549019607843137}\phantom{}100.00 & \cellcolor[rgb]{0.403921568627451,0.713725490196078,0.549019607843137}\phantom{}100.00 & \cellcolor[rgb]{0.403921568627451,0.713725490196078,0.549019607843137}\phantom{}100.00\\ \cellcolor[rgb]{0.40420549201849,0.713707744984139,0.548976230658395}\phantom{0}99.99 & \cellcolor[rgb]{0.403921568627451,0.713725490196078,0.549019607843137}\phantom{}100.00 & \cellcolor[rgb]{0.440677199571511,0.711428263262075,0.543404164226684}\phantom{0}99.18\subcolvspace\\\end{tabular}\subcolend} \\ 
            &  SHRiMP\,2   & \cellcolor[rgb]{0.552754215536619,0.704423449764255,0.52628128678757}\phantom{0}96.53 {\subcolbeg\begin{tabular}{rrr} \cellcolor[rgb]{0.409851866304547,0.71335484659126,0.54811359014247}\phantom{0}99.87 & \cellcolor[rgb]{0.412234814403397,0.713205912335082,0.547749528627368}\phantom{0}99.82 & \cellcolor[rgb]{0.424811998913114,0.712419838303225,0.545828014327272}\phantom{0}99.53\\ \cellcolor[rgb]{0.475953232566866,0.709223511199865,0.538014770296838}\phantom{0}98.37 & \cellcolor[rgb]{0.703774456384545,0.69498468471126,0.503208749991359}\phantom{0}92.58 & \cellcolor[rgb]{0.958415829945575,0.550786109613802,0.478062529591442}\phantom{0}64.63\subcolvspace\\\end{tabular}\subcolend} & \cellcolor[rgb]{0.426188782776781,0.712333789311745,0.545617672348101}\phantom{0}99.50 {\subcolbeg\begin{tabular}{rrr} \cellcolor[rgb]{0.433476422220119,0.711878311846537,0.544504282988702}\phantom{0}99.34 & \cellcolor[rgb]{0.426141925283041,0.712336717905104,0.545624831131867}\phantom{0}99.50 & \cellcolor[rgb]{0.421809884734203,0.712607470439406,0.546286670660161}\phantom{0}99.60\\ \cellcolor[rgb]{0.420005974800686,0.712720214810251,0.546562268011115}\phantom{0}99.64 & \cellcolor[rgb]{0.419714745033467,0.712738416670702,0.546606761447774}\phantom{0}99.65 & \cellcolor[rgb]{0.477794195768349,0.709108450999772,0.537733512029945}\phantom{0}98.32\subcolvspace\\\end{tabular}\subcolend} & \cellcolor[rgb]{0.410618617766177,0.713306924624908,0.547996447558054}\phantom{0}99.85 {\subcolbeg\begin{tabular}{rrr} \cellcolor[rgb]{0.409659018620411,0.713366899571518,0.548143052983102}\phantom{0}99.87 & \cellcolor[rgb]{0.408422332206727,0.713444192472374,0.548331991185192}\phantom{0}99.90 & \cellcolor[rgb]{0.407929583437371,0.713474989270458,0.548407272247177}\phantom{0}99.91\\ \cellcolor[rgb]{0.409032984595831,0.713406026698055,0.54823869707019}\phantom{0}99.89 & \cellcolor[rgb]{0.411356979481178,0.713260777017721,0.54788364229604}\phantom{0}99.84 & \cellcolor[rgb]{0.46724067419355,0.709768046098197,0.539345855603872}\phantom{0}98.57\subcolvspace\\\end{tabular}\subcolend} & \cellcolor[rgb]{0.437462428103346,0.711629186478835,0.543895309867653}\phantom{0}99.25 {\subcolbeg\begin{tabular}{rrr} \cellcolor[rgb]{0.433131218620119,0.711899887071537,0.544557022427591}\phantom{0}99.35 & \cellcolor[rgb]{0.435240635673203,0.711768048505719,0.544234750377814}\phantom{0}99.30 & \cellcolor[rgb]{0.437821709928651,0.711606731364753,0.543840419588787}\phantom{0}99.24\\ \cellcolor[rgb]{0.43511172353298,0.711776105514483,0.544254445288126}\phantom{0}99.30 & \cellcolor[rgb]{0.444285403744607,0.711202750501256,0.54285291081135}\phantom{0}99.09 & \cellcolor[rgb]{0.47085230057144,0.709542319449579,0.538794079351695}\phantom{0}98.48\subcolvspace\\\end{tabular}\subcolend} \\ 
   \bottomrule
\end{tabular}

  }
\end{table*}

\subsubsection{Best-mappers}
Masai showed the best recall values, not loosing more than 3.3\,\% recall on edit distance 5.
Conversely, recall values of BWA and Bowtie\,2 dropped significantly with increasing edit distance up to loosing respectively 15.4\,\% and 11.5\,\% on edit distance 5.
As expected, Soap \,2 turned out to be inadequate for mapping reads of length 100\,bp at this error rates.


\subsubsection{All-mappers}
As expected, RazerS\,3 showed full-sensitivity and mrFAST lost only a minimal percentage of mapping locations.
Overall Masai did not loose more than 0.1\,\% of all mapping locations.
In particular, Masai was full-sensitive for low-error locations and it lost only a small percentage of high-error locations, \ie its loss was limited to 0.1\,\% and 1.4\,\% of mapping locations at edit distance 4 and 5.
These results show that Masai \textcolor{black}{is suited to replace RazerS\,3 or mrFAST as an all-mapper}.

On the other side, BWA and Bowtie\,2 missed 35\,\% and 45\,\% of all mapping locations at edit distance 5 and their recall values as all-mappers did not substantially increase.
Likewise SHRiMP\,2 could not enumerate all mapping locations, although its recall values were good.
Again Hobbes had the worst performance.

\subsection{Variant detection}

The second experiment analyzes the applicability of Masai and other read mappers in \textcolor{black}{genomic} variation pipelines.
Similarly to \cite{Shrimp2}, we simulated from the whole human genome 5 million reads of length $100$\,bp containing sequencing errors, SNPs and indels such that each read had an edit distance of at most 5 to its genomic origin.
To distribute sequencing errors according to a typical Illumina run, we used the read simulator Mason.
The reads were grouped according to the numbers of contained SNPs and indels, where the class $(s,i)$ consists of all reads with $s$ SNPs and $i$ indels.
We mapped the reads with each tool and measured its sensitivity in each class.

We say that a read is mapped \emph{correctly} if a mapping location has been reported within 10\,bp of its genomic origin.
It is considered to map \emph{uniquely} if only one location was reported by the mapper.
For each class we define \emph{recall} to be the fraction of reads which were correctly mapped and \emph{precision} the fraction of uniquely mapped reads that were mapped correctly.
Table~\ref{tab:Variant} shows the results for each read mapper and class.

\begin{table*}[t]
  \caption[Variant detection results]
  {
  \label{tab:Variant}
    {\bfseries Variant detection results.} We show the percentages of found origins (recall) and fraction of unique reads mapped to their origin (precision) classed by reads with $s$ SNPs and $i$ indels $(s,i)$.
  }
  \vspace{-3mm}
  \center
  \sffamily
  \resizebox{1.0\textwidth}{!}
  {
	\renewcommand{\tabcolsep}{0.8ex}
\begin{tabular}{llcccccccccccc}
  \toprule & & \multicolumn{2}{c}{(0,0)} & \multicolumn{2}{c}{(2,0)} & \multicolumn{2}{c}{(4,0)} & \multicolumn{2}{c}{(1,1)} & \multicolumn{2}{c}{(1,2)} & \multicolumn{2}{c}{(0,3)}\\ & method & {  prec.} & {  recl.} & {  prec.} & {  recl.} & {  prec.} & {  recl.} & {  prec.} & {  recl.} & {  prec.} & {  recl.} & {  prec.} & {  recl.} \\ 
  \midrule \multirow{4}{*}{\begin{sideways}\footnotesize best-mappers\hspace{0.3ex} \end{sideways}} & {Masai} & \cellcolor[rgb]{0.483493105779184,0.708752269124095,0.536862845222734}\phantom{0}98.2 & \cellcolor[rgb]{0.484023441732326,0.708719123127024,0.536781821674337}\phantom{0}98.2 & \cellcolor[rgb]{0.509629000004351,0.707118775735022,0.532869861382778}\phantom{0}97.6 & \cellcolor[rgb]{0.513967585872862,0.70684761411824,0.532207021875088}\phantom{0}97.5 & \cellcolor[rgb]{0.539720023075151,0.705238086793097,0.528272621746961}\phantom{0}96.8 & \cellcolor[rgb]{0.539720023075151,0.705238086793097,0.528272621746961}\phantom{0}96.8 & \cellcolor[rgb]{0.499300188712104,0.707764326440788,0.534447874219093}\phantom{0}97.8 & \cellcolor[rgb]{0.525697120480496,0.706114518205263,0.530415009643367}\phantom{0}97.2 & \cellcolor[rgb]{0.494309563102447,0.708076240541391,0.535210330909457}\phantom{0}97.9 & \cellcolor[rgb]{0.494309563102447,0.708076240541391,0.535210330909457}\phantom{0}97.9 & \cellcolor[rgb]{0.524988045587848,0.706158835386054,0.530523340529743}\phantom{0}97.2 & \cellcolor[rgb]{0.524988045587848,0.706158835386054,0.530523340529743}\phantom{0}97.2 \\ 
    & {Bowtie\,2} & \cellcolor[rgb]{0.508838473824568,0.707168183621259,0.5329906362158}\phantom{0}97.6 & \cellcolor[rgb]{0.519048996455205,0.706530025956844,0.531430695258342}\phantom{0}97.3 & \cellcolor[rgb]{0.628476432246483,0.699690811219889,0.51471261479023}\phantom{0}94.6 & \cellcolor[rgb]{0.723843307504668,0.693730381516252,0.500142675514674}\phantom{0}92.0 & \cellcolor[rgb]{0.702754800253591,0.695048413219445,0.503364530789144}\phantom{0}92.6 & \cellcolor[rgb]{0.967489670821669,0.664209120564985,0.4644517682773}\phantom{0}82.5 & \cellcolor[rgb]{0.600488069400511,0.701440083897762,0.518988614669475}\phantom{0}95.3 & \cellcolor[rgb]{0.676726911172774,0.696675156286996,0.507341013843157}\phantom{0}93.3 & \cellcolor[rgb]{0.66882388960936,0.697169095134709,0.508548419915346}\phantom{0}93.5 & \cellcolor[rgb]{0.713352157851341,0.694386078369585,0.501745490045043}\phantom{0}92.3 & \cellcolor[rgb]{0.571639833633641,0.703243098633192,0.523395984022747}\phantom{0}96.1 & \cellcolor[rgb]{0.597184221688353,0.701646574379772,0.519493369181055}\phantom{0}95.4 \\ 
     & {BWA} & \cellcolor[rgb]{0.48397883803087,0.708721910858365,0.536788636128726}\phantom{0}98.2 & \cellcolor[rgb]{0.495154925538074,0.708023405389165,0.535081178315125}\phantom{0}97.9 & \cellcolor[rgb]{0.508597543276224,0.70718324178053,0.533027445049575}\phantom{0}97.6 & \cellcolor[rgb]{0.603399052829862,0.701258147433428,0.518543881089991}\phantom{0}95.3 & \cellcolor[rgb]{0.617743936898315,0.70036159217915,0.516352301579533}\phantom{0}94.9 & \cellcolor[rgb]{0.940986055791384,0.680158959748333,0.46696808897087}\phantom{0}85.1 & \cellcolor[rgb]{0.518402619385443,0.706570424523704,0.531529447310666}\phantom{0}97.4 & \cellcolor[rgb]{0.763555476590209,0.691248370948406,0.494075538571049}\phantom{0}90.9 & \cellcolor[rgb]{0.531321095978248,0.705763019736654,0.529555791164543}\phantom{0}97.1 & \cellcolor[rgb]{0.96598533434621,0.645404914621741,0.466708272990489}\phantom{0}80.3 & \cellcolor[rgb]{0.562898739236122,0.703789417033037,0.524731429000146}\phantom{0}96.3 & \cellcolor[rgb]{0.959082985235469,0.559125550737484,0.4770617966566}\phantom{0}66.5 \\ 
      & {Soap\,2} & \cellcolor[rgb]{0.485445515752829,0.708630243500742,0.536564560365649}\phantom{0}98.1 & \cellcolor[rgb]{0.967766686248491,0.667671813400255,0.464036245137067}\phantom{0}82.9 & \cellcolor[rgb]{0.516182479999456,0.706709183235328,0.531868635272414}\phantom{0}97.4 & \cellcolor[rgb]{0.953231437062921,0.485981198580632,0.485839118915422}\phantom{0}31.0 & \cellcolor[rgb]{0.952941176470588,0.482352941176471,0.486274509803922}\phantom{00}0.0 & \cellcolor[rgb]{0.952941176470588,0.482352941176471,0.486274509803922}\phantom{00}0.0 & \cellcolor[rgb]{0.772425033364672,0.690694023650002,0.492720467397173}\phantom{0}90.6 & \cellcolor[rgb]{0.952941642957199,0.482358772259107,0.486273810074005}\phantom{00}6.2 & \cellcolor[rgb]{0.952941176470588,0.482352941176471,0.486274509803922}\phantom{00}0.0 & \cellcolor[rgb]{0.952941176470588,0.482352941176471,0.486274509803922}\phantom{00}0.0 & \cellcolor[rgb]{0.952941176470588,0.482352941176471,0.486274509803922}\phantom{00}0.0 & \cellcolor[rgb]{0.952941176470588,0.482352941176471,0.486274509803922}\phantom{00}0.0 \\ 
	  \midrule\multirow{5}{*}{\begin{sideways}\footnotesize all-mappers\hspace{0.3ex} \end{sideways}} & {Masai } & \cellcolor[rgb]{0.403931871399493,0.713724846272826,0.54901803380852}\phantom{}100.0 & \cellcolor[rgb]{0.404505893412025,0.713688969897043,0.54893033600105}\phantom{}100.0 & \cellcolor[rgb]{0.403921568627451,0.713725490196078,0.549019607843137}\phantom{}100.0 & \cellcolor[rgb]{0.408708155694196,0.713426328504407,0.54828832370794}\phantom{0}99.9 & \cellcolor[rgb]{0.403921568627451,0.713725490196078,0.549019607843137}\phantom{}100.0 & \cellcolor[rgb]{0.403921568627451,0.713725490196078,0.549019607843137}\phantom{}100.0 & \cellcolor[rgb]{0.403921568627451,0.713725490196078,0.549019607843137}\phantom{}100.0 & \cellcolor[rgb]{0.43534675084361,0.711761416307569,0.544218538337891}\phantom{0}99.3 & \cellcolor[rgb]{0.403921568627451,0.713725490196078,0.549019607843137}\phantom{}100.0 & \cellcolor[rgb]{0.403921568627451,0.713725490196078,0.549019607843137}\phantom{}100.0 & \cellcolor[rgb]{0.403921568627451,0.713725490196078,0.549019607843137}\phantom{}100.0 & \cellcolor[rgb]{0.403921568627451,0.713725490196078,0.549019607843137}\phantom{}100.0 \\ 
       & {RazerS\,3 } & \cellcolor[rgb]{0.403921568627451,0.713725490196078,0.549019607843137}\phantom{}100.0 & \cellcolor[rgb]{0.404238886110199,0.713705657853407,0.548971128783273}\phantom{}100.0 & \cellcolor[rgb]{0.403921568627451,0.713725490196078,0.549019607843137}\phantom{}100.0 & \cellcolor[rgb]{0.403921568627451,0.713725490196078,0.549019607843137}\phantom{}100.0 & \cellcolor[rgb]{0.403921568627451,0.713725490196078,0.549019607843137}\phantom{}100.0 & \cellcolor[rgb]{0.403921568627451,0.713725490196078,0.549019607843137}\phantom{}100.0 & \cellcolor[rgb]{0.403921568627451,0.713725490196078,0.549019607843137}\phantom{}100.0 & \cellcolor[rgb]{0.403921568627451,0.713725490196078,0.549019607843137}\phantom{}100.0 & \cellcolor[rgb]{0.403921568627451,0.713725490196078,0.549019607843137}\phantom{}100.0 & \cellcolor[rgb]{0.403921568627451,0.713725490196078,0.549019607843137}\phantom{}100.0 & \cellcolor[rgb]{0.403921568627451,0.713725490196078,0.549019607843137}\phantom{}100.0 & \cellcolor[rgb]{0.403921568627451,0.713725490196078,0.549019607843137}\phantom{}100.0 \\ 
        & {Hobbes } & \cellcolor[rgb]{0.406517422875985,0.713563249305545,0.548623018999611}\phantom{0}99.9 & \cellcolor[rgb]{0.40631556422282,0.713575865471368,0.548653858516067}\phantom{0}99.9 & \cellcolor[rgb]{0.406220541341242,0.713581804401466,0.548668375900753}\phantom{0}99.9 & \cellcolor[rgb]{0.406615899857705,0.713557094494188,0.548607973905182}\phantom{0}99.9 & \cellcolor[rgb]{0.403921568627451,0.713725490196078,0.549019607843137}\phantom{}100.0 & \cellcolor[rgb]{0.403921568627451,0.713725490196078,0.549019607843137}\phantom{}100.0 & \cellcolor[rgb]{0.403921568627451,0.713725490196078,0.549019607843137}\phantom{}100.0 & \cellcolor[rgb]{0.411840288430082,0.713230570208414,0.547809803428846}\phantom{0}99.8 & \cellcolor[rgb]{0.403921568627451,0.713725490196078,0.549019607843137}\phantom{}100.0 & \cellcolor[rgb]{0.665681170482356,0.697365515080147,0.509028557559749}\phantom{0}93.6 & \cellcolor[rgb]{0.419898848635959,0.712726910195547,0.546578634508504}\phantom{0}99.6 & \cellcolor[rgb]{0.775805943719934,0.690482716752798,0.49220393942623}\phantom{0}90.5 \\ 
         & {mrFAST } & \cellcolor[rgb]{0.40588758362824,0.713602614258529,0.548719244440239}\phantom{}100.0 & \cellcolor[rgb]{0.408921870735986,0.713412971314295,0.548255672798778}\phantom{0}99.9 & \cellcolor[rgb]{0.404081754676621,0.713715478568005,0.548995134974514}\phantom{}100.0 & \cellcolor[rgb]{0.404370959048354,0.713697403294772,0.548950950973277}\phantom{}100.0 & \cellcolor[rgb]{0.403921568627451,0.713725490196078,0.549019607843137}\phantom{}100.0 & \cellcolor[rgb]{0.403921568627451,0.713725490196078,0.549019607843137}\phantom{}100.0 & \cellcolor[rgb]{0.403921568627451,0.713725490196078,0.549019607843137}\phantom{}100.0 & \cellcolor[rgb]{0.403921568627451,0.713725490196078,0.549019607843137}\phantom{}100.0 & \cellcolor[rgb]{0.403921568627451,0.713725490196078,0.549019607843137}\phantom{}100.0 & \cellcolor[rgb]{0.403921568627451,0.713725490196078,0.549019607843137}\phantom{}100.0 & \cellcolor[rgb]{0.403921568627451,0.713725490196078,0.549019607843137}\phantom{}100.0 & \cellcolor[rgb]{0.403921568627451,0.713725490196078,0.549019607843137}\phantom{}100.0 \\ 
          & {SHRiMP\,2 } & \cellcolor[rgb]{0.403990550529113,0.713721178827225,0.549009068941494}\phantom{}100.0 & \cellcolor[rgb]{0.429550840032526,0.712123660733261,0.545104024711806}\phantom{0}99.4 & \cellcolor[rgb]{0.404131677087168,0.713712358417346,0.548987507939569}\phantom{}100.0 & \cellcolor[rgb]{0.41823566990134,0.71283085886646,0.546832731259627}\phantom{0}99.7 & \cellcolor[rgb]{0.403921568627451,0.713725490196078,0.549019607843137}\phantom{}100.0 & \cellcolor[rgb]{0.416810586683264,0.71291992656759,0.547050452306833}\phantom{0}99.7 & \cellcolor[rgb]{0.403921568627451,0.713725490196078,0.549019607843137}\phantom{}100.0 & \cellcolor[rgb]{0.424945529566915,0.712411492637362,0.545807613810719}\phantom{0}99.5 & \cellcolor[rgb]{0.403921568627451,0.713725490196078,0.549019607843137}\phantom{}100.0 & \cellcolor[rgb]{0.437445363772424,0.711630252999518,0.543897916918211}\phantom{0}99.2 & \cellcolor[rgb]{0.403921568627451,0.713725490196078,0.549019607843137}\phantom{}100.0 & \cellcolor[rgb]{0.423988352149138,0.712471316225973,0.545953849249546}\phantom{0}99.6 \\ 
   \bottomrule \end{tabular}

  }
\end{table*}

\subsubsection{Best-mappers}
Among best-mappers, Masai showed the highest precision and recall in all classes.
In particular, Masai did not loose more than 3.2\,\% recall in class (4,0), whether Bowtie\,2 and BWA lost respectively 17.5\,\% and 14.9\,\% and Soap\,2 was not able to map any read.

Interestingly, we observed that recall values of Bowtie\,2, BWA and Soap\,2 were negatively correlated with the amount of \textcolor{black}{genomic} variation.
For instance, in the Rabema benchmark Bowtie\,2 lost respectively 7.2\,\% and 11.5\,\% of mapping locations at distance 4 and 5, but in this experiment it lost 17.5\,\% recall in class (4,0).
We noticed a similar trend for BWA and Soap\,2.
These tools rely on quality values to guess the best mapping location for a read and tend to prefer alignments which can be explained by sequencing errors instead of true \textcolor{black}{genomic} variations.
The low performance of Soap\,2 is also due to its limitation to at most 2 mismatches and no support for indels.

\subsubsection{All-mappers}
Looking at all-mappers results, Masai showed 100\,\% precision and recall in all classes, except for classes (2,0) and (1,1) where it lost only 0.1\,\% and 0.7\,\% recall.
Masai is therefore roughly comparable to the full-sensitive read mappers RazerS\,3 and mrFAST.
SHRiMP\,2 showed 100\,\% precision in all classes but lost between 0.3\,\% and 0.8\,\% recall in each class.
Hobbes had the lowest performance among all-mappers.
It appears to have problems with indels, indeed it lost 9.5\,\% recall in class (0,3).

\subsection{Runtime \textcolor{black}{on real data}}

In the last experiment we compared the runtime of Masai with the other read mappers.
To this end, we mapped the first $10\,\text{M}\times 100\,\text{bp}$ reads from an Illumina lane of E.~coli, C.~elegans, D.~melanogaster and H.~sapiens.
Whenever possible we asked mappers to map reads within edit distance 5.
We measured running times, peak memory consumptions, mapped reads and Rabema any-best scores.

For the evaluation we adopted the commonly used measure of percentage of \emph{mapped reads}, \ie the fraction of reads for which the read mapper reports a mapping location.
However, as some mappers report mapping locations without constraints on the number of errors, we also included Rabema \emph{any-best} scores.
The Rabema any-best benchmark assigns a point for a read if the mapper reports at least one mapping location at the minimum edit distance.
Final Rabema any-best scores are normalized by the number of reads.

Results for C.~elegans and H.~sapiens are shown in Table~\ref{tab:Runtime}.
Additional results for E.~coli and D.~melanogaster are shown in \textcolor{black}{the Supplementary Data}.

\begin{landscape}
\begin{table*}[t]
  \caption[Runtime results]{
    \label{tab:Runtime}
    \textbf{Runtime results.}
	Results of mapping $10\,\text{M}\times 100\,\text{bp}$ Illumina reads.
	\textbf{Mapped reads.}
	In large we show the percentage of mapped reads and in small the cumulative percentage of reads that were mapped with $\bigl(\begin{smallmatrix}\mbox{\tiny 0}&\mbox{\tiny 1\%}&\mbox{\tiny 2\%}\\\mbox{\tiny 3\%}&\mbox{\tiny 4\%}&\mbox{\tiny 5\%}\end{smallmatrix}\bigr)$ errors.
	\textbf{Rabema any-best.}
    In large we show the percentage of reads mapped with the minimal number of errors (up to 5\%) and in small the percentage of reads that were mapped with $\bigl(\begin{smallmatrix}\mbox{\tiny 0}&\mbox{\tiny 1\%}&\mbox{\tiny 2\%}\\\mbox{\tiny 3\%}&\mbox{\tiny 4\%}&\mbox{\tiny 5\%}\end{smallmatrix}\bigr)$ errors.
	\textbf{Remarks.}
    SHRiMP\,2 was not able to map the H.~sapiens dataset within 4 days.
    Hobbes constantly crashed and was not able to map completely nor the C.~Elegans nor the H.~sapiens dataset.
  }
	\vspace{-3mm}
	\center
	\sffamily
	\resizebox{1.4\textwidth}{!}
	{
		\renewcommand{\tabcolsep}{0.8ex}
\begin{tabular}{llrrccrrcc}
  \toprule
  & \multirow{2}{*}{\quad dataset}  &\multicolumn{ 4 }{c}{  SRR065390 } &\multicolumn{ 4 }{c}{  ERR012100 } \\
  &&\multicolumn{4}{c}{C.\,elegans}&\multicolumn{4}{c}{H.\,sapiens} \\
  \cmidrule(lr){3-6}\cmidrule(lr){7-10} 
  &  &\multicolumn{1}{c}{  time } &\multicolumn{1}{c}{  memory } &\multicolumn{1}{c}{  Rabema any-best } &\multicolumn{1}{c}{  mapped reads } &\multicolumn{1}{c}{  time } &\multicolumn{1}{c}{  memory } &\multicolumn{1}{c}{  Rabema any-best } &\multicolumn{1}{c}{  mapped reads } \\
  & method  &\multicolumn{1}{c}{  [min:s] } &\multicolumn{1}{c}{  [Mb] } &\multicolumn{1}{c}{  [\%] } &\multicolumn{1}{c}{  [\%] } &\multicolumn{1}{c}{  [min:s] } &\multicolumn{1}{c}{  [Mb] } &\multicolumn{1}{c}{  [\%] } &\multicolumn{1}{c}{  [\%] } \\
  \midrule
\multirow{4}{*}{\begin{sideways}\footnotesize best-mappers \hspace{1.1ex} \end{sideways}} &  Masai  & \phantom{000}3:10\ \  & 6006\ \ \  & \cellcolor[rgb]{0.403921568627451,0.713725490196078,0.549019607843137}\phantom{}100.00 \subcolbeg\begin{tabular}{rrr} \cellcolor[rgb]{0.403921568627451,0.713725490196078,0.549019607843137}\phantom{}100.00 & \cellcolor[rgb]{0.403921568627451,0.713725490196078,0.549019607843137}\phantom{}100.00 & \cellcolor[rgb]{0.403921568627451,0.713725490196078,0.549019607843137}\phantom{}100.00 \\ \cellcolor[rgb]{0.403921568627451,0.713725490196078,0.549019607843137}\phantom{}100.00 & \cellcolor[rgb]{0.403921568627451,0.713725490196078,0.549019607843137}\phantom{}100.00 & \cellcolor[rgb]{0.403921568627451,0.713725490196078,0.549019607843137}\phantom{}100.00\subcolvspace \\ \end{tabular}\subcolend & \phantom{0}89.49 \subcolbeg\begin{tabular}{rrr} \phantom{0}75.01 & \phantom{0}83.80 & \phantom{0}86.38 \\ \phantom{0}87.83 & \phantom{0}88.79 & \phantom{0}89.49\subcolvspace \\ \end{tabular}\subcolend & \phantom{00}24:56\ \  & 44736\ \ \  & \cellcolor[rgb]{0.404178823033906,0.713709411795675,0.548980305086596}\phantom{0}99.99 \subcolbeg\begin{tabular}{rrr} \cellcolor[rgb]{0.403921568627451,0.713725490196078,0.549019607843137}\phantom{}100.00 & \cellcolor[rgb]{0.403921568627451,0.713725490196078,0.549019607843137}\phantom{}100.00 & \cellcolor[rgb]{0.403921568627451,0.713725490196078,0.549019607843137}\phantom{}100.00 \\ \cellcolor[rgb]{0.406550573370574,0.713561177399633,0.548617954340716}\phantom{0}99.94 & \cellcolor[rgb]{0.40802639599099,0.713468938485857,0.548392481440374}\phantom{0}99.91 & \cellcolor[rgb]{0.425102005713463,0.712401712878203,0.545783707732774}\phantom{0}99.53\subcolvspace \\ \end{tabular}\subcolend & \phantom{0}93.76 \subcolbeg\begin{tabular}{rrr} \phantom{0}75.99 & \phantom{0}87.84 & \phantom{0}90.67 \\ \phantom{0}92.02 & \phantom{0}92.99 & \phantom{0}93.76\subcolvspace \\ \end{tabular}\subcolend \\ 
    &  Bowtie\,2  & \phantom{00}24:14\ \  & 135\ \ \  & \cellcolor[rgb]{0.439104893175053,0.711526532411853,0.54364437770392}\phantom{0}99.21 \subcolbeg\begin{tabular}{rrr} \cellcolor[rgb]{0.403921568627451,0.713725490196078,0.549019607843137}\phantom{}100.00 & \cellcolor[rgb]{0.435206888901783,0.711770157678933,0.544239906134559}\phantom{0}99.30 & \cellcolor[rgb]{0.674493366057199,0.696814752856719,0.507682249902481}\phantom{0}93.38 \\ \cellcolor[rgb]{0.83704882497921,0.686655036674093,0.48284738812273}\phantom{0}88.61 & \cellcolor[rgb]{0.968579505493531,0.677832053963251,0.462817016269508}\phantom{0}84.03 & \cellcolor[rgb]{0.96452748809715,0.627181836508491,0.468895042364079}\phantom{0}77.96\subcolvspace \\ \end{tabular}\subcolend & \phantom{0}92.58 \subcolbeg\begin{tabular}{rrr} \phantom{0}75.01 & \phantom{0}83.74 & \phantom{0}86.20 \\ \phantom{0}87.61 & \phantom{0}88.57 & \phantom{0}89.27\subcolvspace \\ \end{tabular}\subcolend & \phantom{00}57:41\ \  & 3180\ \ \  & \cellcolor[rgb]{0.428675142534545,0.712178391826885,0.545237811829554}\phantom{0}99.45 \subcolbeg\begin{tabular}{rrr} \cellcolor[rgb]{0.403921568627451,0.713725490196078,0.549019607843137}\phantom{}100.00 & \cellcolor[rgb]{0.415391463682881,0.713008621755114,0.547267262765224}\phantom{0}99.75 & \cellcolor[rgb]{0.57346114789461,0.703129266491881,0.523117727677321}\phantom{0}96.02 \\ \cellcolor[rgb]{0.692742897622678,0.695674157133877,0.504894127024422}\phantom{0}92.88 & \cellcolor[rgb]{0.860438980108602,0.685193151978506,0.479273892200184}\phantom{0}87.86 & \cellcolor[rgb]{0.965324378974874,0.637142972480045,0.467699706047493}\phantom{0}79.26\subcolvspace \\ \end{tabular}\subcolend & \phantom{0}96.72 \subcolbeg\begin{tabular}{rrr} \phantom{0}75.99 & \phantom{0}87.81 & \phantom{0}90.54 \\ \phantom{0}91.85 & \phantom{0}92.76 & \phantom{0}93.44\subcolvspace \\ \end{tabular}\subcolend \\ 
     &  BWA  & \phantom{00}25:53\ \  & 325\ \ \  & \cellcolor[rgb]{0.43408514443473,0.711840266708124,0.54441128376147}\phantom{0}99.33 \subcolbeg\begin{tabular}{rrr} \cellcolor[rgb]{0.403922773160041,0.713725414912792,0.549019423817325}\phantom{}100.00 & \cellcolor[rgb]{0.444566434173703,0.711185186099438,0.542809975606904}\phantom{0}99.09 & \cellcolor[rgb]{0.591120724887586,0.70202554292982,0.520419736747839}\phantom{0}95.57 \\ \cellcolor[rgb]{0.80201766608537,0.688844484104959,0.488199370731511}\phantom{0}89.70 & \cellcolor[rgb]{0.919667482150992,0.681491370600857,0.470225093277041}\phantom{0}85.86 & \cellcolor[rgb]{0.96732533876891,0.662154969905493,0.464698266356439}\phantom{0}82.29\subcolvspace \\ \end{tabular}\subcolend & \phantom{0}89.33 \subcolbeg\begin{tabular}{rrr} \phantom{0}75.01 & \phantom{0}83.72 & \phantom{0}86.25 \\ \phantom{0}87.64 & \phantom{0}88.59 & \phantom{0}89.32\subcolvspace \\ \end{tabular}\subcolend & \phantom{00}80:58\ \  & 4475\ \ \  & \cellcolor[rgb]{0.424675171731692,0.712428390002063,0.545848918479989}\phantom{0}99.54 \subcolbeg\begin{tabular}{rrr} \cellcolor[rgb]{0.403921568627451,0.713725490196078,0.549019607843137}\phantom{}100.00 & \cellcolor[rgb]{0.426492250046142,0.71231482260741,0.545571309293059}\phantom{0}99.50 & \cellcolor[rgb]{0.490989571590254,0.708283740010903,0.535717551834931}\phantom{0}98.01 \\ \cellcolor[rgb]{0.67436508922478,0.696822770158745,0.507701847751879}\phantom{0}93.39 & \cellcolor[rgb]{0.827280690985052,0.687265545048728,0.484339741927393}\phantom{0}88.92 & \cellcolor[rgb]{0.959571126227324,0.678997392846086,0.464128703209823}\phantom{0}84.42\subcolvspace \\ \end{tabular}\subcolend & \phantom{0}93.53 \subcolbeg\begin{tabular}{rrr} \phantom{0}75.99 & \phantom{0}87.78 & \phantom{0}90.59 \\ \phantom{0}91.91 & \phantom{0}92.82 & \phantom{0}93.53\subcolvspace \\ \end{tabular}\subcolend \\ 
      &  Soap\,2  & \phantom{000}4:37\ \  & 748\ \ \  & \cellcolor[rgb]{0.575011331558313,0.7030323800129,0.522880894062033}\phantom{0}95.98 \subcolbeg\begin{tabular}{rrr} \cellcolor[rgb]{0.403921568627451,0.713725490196078,0.549019607843137}\phantom{}100.00 & \cellcolor[rgb]{0.551032821733812,0.704531036876931,0.526544277507443}\phantom{0}96.57 & \cellcolor[rgb]{0.710643807160784,0.694555350287745,0.502159265844989}\phantom{0}92.38 \\ \cellcolor[rgb]{0.952941176474304,0.482352941222915,0.486274509798348}\phantom{00}0.33 & \cellcolor[rgb]{0.952941176470589,0.482352941176478,0.486274509803921}\phantom{00}0.04 & \cellcolor[rgb]{0.952941176470588,0.482352941176471,0.486274509803922}\phantom{00}0.02\subcolvspace \\ \end{tabular}\subcolend & \phantom{0}85.95 \subcolbeg\begin{tabular}{rrr} \phantom{0}75.01 & \phantom{0}83.50 & \phantom{0}85.94 \\ \phantom{0}85.95 & \phantom{0}85.95 & \phantom{0}85.95\subcolvspace \\ \end{tabular}\subcolend & \phantom{00}11:11\ \  & 5357\ \ \  & \cellcolor[rgb]{0.587778558209632,0.702234428347192,0.52093034554586}\phantom{0}95.66 \subcolbeg\begin{tabular}{rrr} \cellcolor[rgb]{0.403924541315772,0.713725304403058,0.549019153682421}\phantom{}100.00 & \cellcolor[rgb]{0.615757824038359,0.700485724232897,0.516655735488693}\phantom{0}94.94 & \cellcolor[rgb]{0.899747537431236,0.682736367145842,0.473268418164781}\phantom{0}86.54 \\ \cellcolor[rgb]{0.952941176473969,0.482352941218729,0.486274509798851}\phantom{00}0.32 & \cellcolor[rgb]{0.952941176470795,0.482352941179057,0.486274509803611}\phantom{00}0.16 & \cellcolor[rgb]{0.952941176470779,0.482352941178851,0.486274509803636}\phantom{00}0.16\subcolvspace \\ \end{tabular}\subcolend & \phantom{0}89.73 \subcolbeg\begin{tabular}{rrr} \phantom{0}75.99 & \phantom{0}87.24 & \phantom{0}89.73 \\ \phantom{0}89.73 & \phantom{0}89.73 & \phantom{0}89.73\subcolvspace \\ \end{tabular}\subcolend \\ 
  \midrule\multirow{5}{*}{\begin{sideways}\footnotesize all-mappers \hspace{2.4ex} \end{sideways}} &  Masai   & \phantom{00}10:48\ \  & 6006\ \ \  & \cellcolor[rgb]{0.403921568627451,0.713725490196078,0.549019607843137}\phantom{}100.00 \subcolbeg\begin{tabular}{rrr} \cellcolor[rgb]{0.403921568627451,0.713725490196078,0.549019607843137}\phantom{}100.00 & \cellcolor[rgb]{0.403921568627451,0.713725490196078,0.549019607843137}\phantom{}100.00 & \cellcolor[rgb]{0.403921568627451,0.713725490196078,0.549019607843137}\phantom{}100.00 \\ \cellcolor[rgb]{0.403921568627451,0.713725490196078,0.549019607843137}\phantom{}100.00 & \cellcolor[rgb]{0.403921568627451,0.713725490196078,0.549019607843137}\phantom{}100.00 & \cellcolor[rgb]{0.403921568627451,0.713725490196078,0.549019607843137}\phantom{}100.00\subcolvspace \\ \end{tabular}\subcolend & \phantom{0}89.49 \subcolbeg\begin{tabular}{rrr} \phantom{0}75.01 & \phantom{0}83.80 & \phantom{0}86.38 \\ \phantom{0}87.83 & \phantom{0}88.79 & \phantom{0}89.49\subcolvspace \\ \end{tabular}\subcolend & \phantom{0}284:34\ \  & 57319\ \ \  & \cellcolor[rgb]{0.404113309229204,0.713713506408469,0.548990314140092}\phantom{}100.00 \subcolbeg\begin{tabular}{rrr} \cellcolor[rgb]{0.403921568627451,0.713725490196078,0.549019607843137}\phantom{}100.00 & \cellcolor[rgb]{0.403921568627451,0.713725490196078,0.549019607843137}\phantom{}100.00 & \cellcolor[rgb]{0.403921568627451,0.713725490196078,0.549019607843137}\phantom{}100.00 \\ \cellcolor[rgb]{0.404021488201674,0.713719245222689,0.549004342352631}\phantom{}100.00 & \cellcolor[rgb]{0.40522886610987,0.713643784103427,0.548819881838879}\phantom{0}99.97 & \cellcolor[rgb]{0.425102005713463,0.712401712878203,0.545783707732774}\phantom{0}99.53\subcolvspace \\ \end{tabular}\subcolend & \phantom{0}93.76 \subcolbeg\begin{tabular}{rrr} \phantom{0}75.99 & \phantom{0}87.84 & \phantom{0}90.67 \\ \phantom{0}92.02 & \phantom{0}92.99 & \phantom{0}93.76\subcolvspace \\ \end{tabular}\subcolend \\ 
       &  RazerS\,3   & \phantom{00}21:18\ \  & 11489\ \ \  & \cellcolor[rgb]{0.403921568627451,0.713725490196078,0.549019607843137}\phantom{}100.00 \subcolbeg\begin{tabular}{rrr} \cellcolor[rgb]{0.403921568627451,0.713725490196078,0.549019607843137}\phantom{}100.00 & \cellcolor[rgb]{0.403921568627451,0.713725490196078,0.549019607843137}\phantom{}100.00 & \cellcolor[rgb]{0.403921568627451,0.713725490196078,0.549019607843137}\phantom{}100.00 \\ \cellcolor[rgb]{0.403921568627451,0.713725490196078,0.549019607843137}\phantom{}100.00 & \cellcolor[rgb]{0.403921568627451,0.713725490196078,0.549019607843137}\phantom{}100.00 & \cellcolor[rgb]{0.403921568627451,0.713725490196078,0.549019607843137}\phantom{}100.00\subcolvspace \\ \end{tabular}\subcolend & \phantom{0}89.49 \subcolbeg\begin{tabular}{rrr} \phantom{0}75.01 & \phantom{0}83.80 & \phantom{0}86.38 \\ \phantom{0}87.83 & \phantom{0}88.79 & \phantom{0}89.49\subcolvspace \\ \end{tabular}\subcolend & \phantom{}3653:03\ \  & 17298\ \ \  & \cellcolor[rgb]{0.403921568627451,0.713725490196078,0.549019607843137}\phantom{}100.00 \subcolbeg\begin{tabular}{rrr} \cellcolor[rgb]{0.403921568627451,0.713725490196078,0.549019607843137}\phantom{}100.00 & \cellcolor[rgb]{0.403921568627451,0.713725490196078,0.549019607843137}\phantom{}100.00 & \cellcolor[rgb]{0.403921568627451,0.713725490196078,0.549019607843137}\phantom{}100.00 \\ \cellcolor[rgb]{0.403921568627451,0.713725490196078,0.549019607843137}\phantom{}100.00 & \cellcolor[rgb]{0.403921568627451,0.713725490196078,0.549019607843137}\phantom{}100.00 & \cellcolor[rgb]{0.403921568627451,0.713725490196078,0.549019607843137}\phantom{}100.00\subcolvspace \\ \end{tabular}\subcolend & \phantom{0}93.77 \subcolbeg\begin{tabular}{rrr} \phantom{0}75.99 & \phantom{0}87.84 & \phantom{0}90.67 \\ \phantom{0}92.02 & \phantom{0}92.99 & \phantom{0}93.77\subcolvspace \\ \end{tabular}\subcolend \\ 
        &  Hobbes   & \phantom{0}117:46\ \  & 3885\ \ \  & \cellcolor[rgb]{0.799874862228782,0.688978409345995,0.488526743542934}\phantom{0}89.77 \subcolbeg\begin{tabular}{rrr} \cellcolor[rgb]{0.75739865651706,0.691633172202978,0.495016163860003}\phantom{0}91.04 & \cellcolor[rgb]{0.966201859919643,0.64811148428966,0.466383484630339}\phantom{0}80.63 & \cellcolor[rgb]{0.902035590281386,0.682593363842708,0.472918854534897}\phantom{0}86.47 \\ \cellcolor[rgb]{0.846069273303926,0.686091258653799,0.48146926407312}\phantom{0}88.32 & \cellcolor[rgb]{0.841457767812344,0.686379477747023,0.482173799634334}\phantom{0}88.47 & \cellcolor[rgb]{0.939135451406109,0.680274622522412,0.467250820196398}\phantom{0}85.17\subcolvspace \\ \end{tabular}\subcolend & \phantom{0}80.34 \subcolbeg\begin{tabular}{rrr} \phantom{0}68.29 & \phantom{0}75.38 & \phantom{0}77.61 \\ \phantom{0}78.89 & \phantom{0}79.74 & \phantom{0}80.34\subcolvspace \\ \end{tabular}\subcolend & \phantom{}2319:27\ \  & 71685\ \ \  & \cellcolor[rgb]{0.956749046432414,0.52995131569929,0.480562704861183}\phantom{0}59.02 \subcolbeg\begin{tabular}{rrr} \cellcolor[rgb]{0.956803646581621,0.530633817564379,0.480480804637373}\phantom{0}59.24 & \cellcolor[rgb]{0.956653638990009,0.528758722669231,0.48070581602479}\phantom{0}58.65 & \cellcolor[rgb]{0.956366643340757,0.525171277053586,0.481136309498668}\phantom{0}57.48 \\ \cellcolor[rgb]{0.956233820101892,0.523510986567769,0.481335544356966}\phantom{0}56.92 & \cellcolor[rgb]{0.956182957289943,0.522875201418401,0.48141183857489}\phantom{0}56.70 & \cellcolor[rgb]{0.956098132399845,0.521814890292186,0.481539075910036}\phantom{0}56.32\subcolvspace \\ \end{tabular}\subcolend & \phantom{0}55.35 \subcolbeg\begin{tabular}{rrr} \phantom{0}45.01 & \phantom{0}51.96 & \phantom{0}53.59 \\ \phantom{0}54.36 & \phantom{0}54.91 & \phantom{0}55.35\subcolvspace \\ \end{tabular}\subcolend \\ 
         &  mrFAST   & \phantom{00}67:41\ \  & 875\ \ \  & \cellcolor[rgb]{0.404460115873293,0.713691830993213,0.548937329791689}\phantom{0}99.99 \subcolbeg\begin{tabular}{rrr} \cellcolor[rgb]{0.403921568627451,0.713725490196078,0.549019607843137}\phantom{}100.00 & \cellcolor[rgb]{0.404717916288039,0.713675718467292,0.548897943617214}\phantom{0}99.98 & \cellcolor[rgb]{0.409408296230922,0.713382569720862,0.548181357792607}\phantom{0}99.88 \\ \cellcolor[rgb]{0.409566063127787,0.713372709289807,0.548157254516697}\phantom{0}99.87 & \cellcolor[rgb]{0.407267255996864,0.71351638473549,0.548508461161699}\phantom{0}99.93 & \cellcolor[rgb]{0.425950445603491,0.712348685385076,0.545654084971798}\phantom{0}99.51\subcolvspace \\ \end{tabular}\subcolend & \phantom{0}89.49 \subcolbeg\begin{tabular}{rrr} \phantom{0}75.01 & \phantom{0}83.80 & \phantom{0}86.38 \\ \phantom{0}87.83 & \phantom{0}88.78 & \phantom{0}89.49\subcolvspace \\ \end{tabular}\subcolend & \phantom{}4462:25\ \  & 929\ \ \  & \cellcolor[rgb]{0.404881953443559,0.713665466145072,0.548872882385121}\phantom{0}99.98 \subcolbeg\begin{tabular}{rrr} \cellcolor[rgb]{0.403921568627451,0.713725490196078,0.549019607843137}\phantom{}100.00 & \cellcolor[rgb]{0.403921568627451,0.713725490196078,0.549019607843137}\phantom{}100.00 & \cellcolor[rgb]{0.403953520074563,0.713723493230634,0.549014726372051}\phantom{}100.00 \\ \cellcolor[rgb]{0.40418800609198,0.713708837854545,0.54897890211939}\phantom{0}99.99 & \cellcolor[rgb]{0.404201799402959,0.713707975772609,0.54897679480799}\phantom{0}99.99 & \cellcolor[rgb]{0.514573728445897,0.706809730207426,0.532114416759764}\phantom{0}97.46\subcolvspace \\ \end{tabular}\subcolend & \phantom{0}93.75 \subcolbeg\begin{tabular}{rrr} \phantom{0}75.99 & \phantom{0}87.84 & \phantom{0}90.67 \\ \phantom{0}92.02 & \phantom{0}92.99 & \phantom{0}93.75\subcolvspace \\ \end{tabular}\subcolend \\ 
          &  SHRiMP\,2   & \phantom{0}541:20\ \  & 2735\ \ \  & \cellcolor[rgb]{0.469556092795004,0.709623332435606,0.538992111095317}\phantom{0}98.51 \subcolbeg\begin{tabular}{rrr} \cellcolor[rgb]{0.422209307876188,0.712582506493032,0.546225647680136}\phantom{0}99.59 & \cellcolor[rgb]{0.541444929765213,0.705130280124968,0.528009094335979}\phantom{0}96.81 & \cellcolor[rgb]{0.732750268086859,0.693173696479865,0.498781889870172}\phantom{0}91.76 \\ \cellcolor[rgb]{0.868184440322121,0.684709060715162,0.478090558000896}\phantom{0}87.60 & \cellcolor[rgb]{0.967042135962828,0.658614934829471,0.465123070565562}\phantom{0}81.88 & \cellcolor[rgb]{0.962747932951192,0.604937397184014,0.471564375083016}\phantom{0}74.77\subcolvspace \\ \end{tabular}\subcolend & \phantom{0}91.91 \subcolbeg\begin{tabular}{rrr} \phantom{0}74.71 & \phantom{0}83.22 & \phantom{0}85.59 \\ \phantom{0}86.88 & \phantom{0}87.69 & \phantom{0}88.27\subcolvspace \\ \end{tabular}\subcolend & --\ \  & --\ \  & -- & -- \\ 
   \bottomrule
\end{tabular}

	}
\end{table*}
\end{landscape}

\subsubsection{Best-mappers}
On the C.~elegans dataset Masai was 7.7 times faster than Bowtie\,2, 8.2 times faster than BWA and 1.5 times faster than Soap\,2.
On the H.~sapiens dataset Masai was 2.3 times faster than Bowtie\,2, 3.2 times faster than BWA but 2.2 times slower than Soap\,2.
On one hand, Soap\,2 was not able to map a consistent fraction of reads because of its limitation to 2 mismatches.
On the other hand, Bowtie\,2 reported more mapped reads than Masai but, taking any-best scores into account, it reported less mapping locations than Masai.
\textcolor{black}{On the C.~elegans and H.~sapiens datasets, Bowtie\,2 missed respectively 22.0\,\% and 20.7\,\% of reads mappable at edit distance 5.}
This is due to the fact that Bowtie\,2 uses a scoring scheme based on quality values and does not impose a maximal error rate threshold.


\subsubsection{All-mappers}
On the C.~elegans dataset Masai was 2.0 times faster than RazerS\,3, 10.9 times faster than Hobbes, 6.3 times faster than mrFAST and 50.1 times faster than SHRiMP\,2.
Hobbes constantly crashed and mapped less reads than all other mappers in this category.
Likewise for Bowtie\,2, also SHRiMP\,2 does not impose a maximal error rate threshold and reported more mapped reads than Masai.
However its Rabema any-best score was inferior to Masai.
This could be due to the use of a different scoring scheme where two mismatches cost less than opening a gap.
Anyway this hypothesis does not explain why SHRiMP\,2 did not report some mapping locations at distance 0.

On the H.~sapiens dataset Masai was 12.8 times faster than RazerS\,3, 15.7 times faster than mrFAST, \textcolor{black}{and 8.2 times faster than Hobbes which however constantly crashed and mapped only half of the reads.}
SHRiMP\,2 was not able to map the H.~sapiens dataset within 4 days.


\section{Discussion}

\textcolor{black}{
We showed that, on one hand Masai is faster and more accurate than the best-mappers Bowtie\,2 and BWA, while on the other hand Masai is slightly slower but substantially more accurate than Soap\,2.}
Masai's accuracy becomes considerable in presence of genomic variation, therefore we strongly advise to use Masai in small and large genomic variation pipelines.

At the same time, we showed that Masai is \textcolor{black}{significantly} faster than any other all-mapper while being almost full-sensitive. Consequently Masai brings all-mapping within feasible times, although with a higher memory footprint.

In the near future, we plan to index reference genomes using the suffix array or the FM-index \textcolor{black}{to reduce the memory consumption.
To achieve full-sensitive mapping on edit distance we will extend multiple backtracking to consider indels.}

Masai is implemented in \CC using the SeqAn library.
The source code is distributed under the BSD license and binaries for Linux, Mac OS X and Windows can be freely downloaded from \url{http://www.seqan.de/projects/masai}.


\section{Acknowledgements}
We thank Manuel Holtgrewe for his joint work on experimental evaluations.



\bibliographystyle{nar}
\bibliography{masai}


\end{document}